\documentclass[final]{jfm}

\usepackage{graphics}
\usepackage{xcolor} 
\usepackage{graphicx}
\usepackage{subcaption} 

\usepackage{newtxtext}
\usepackage{newtxmath}
\usepackage{natbib}
\usepackage{hyperref}
\hypersetup{
    colorlinks = true,
    urlcolor   = blue,
    citecolor  = black,
}

\newcommand{\RomanNumeralCaps}[1]
\linenumbers

\newcommand{\Rag}{\operatorname{\mathit{Ra_g}}}
\newcommand{\Ram}{\operatorname{\mathit{Ra_m}}}

\title{How gravity stabilises instability: the case of magnetic micro-convection}

\author{Andrejs Tatu\c l\v cenkovs\aff{1}
  \corresp{\email{andrejs.tatulcenkovs@lu.lv}},
  L\=asma Pu\c kina-Slava\aff{1},
  Andrejs C\=ebers\aff{1}
 \and Guntars Kitenbergs\aff{1}
  \corresp{\email{guntars.kitenbergs@lu.lv}}
  }

\affiliation{\aff{1}MMML lab, Department of Physics, University of Latvia}

\begin{document}
\maketitle

\begin{abstract}
Finding solutions for better mixing in microfluidics remains an important challenge, including understanding fundamental aspects of these processes.
Here we investigate the magnetic micro-convection on water and miscible magnetic fluid interface in a vertical microfluidic chip to understand what is the role of gravity, as fluids have different densities.
Our model is reduced to two dimensionless quantities - magnetic and gravitational Rayleigh numbers.
Numerical simulation results show that static magnetic field generate rich dynamics.
This is confirmed quantitatively with careful experiments in initially stagnant fluids.
We also show that the length of resulting mixing is limited by gravity.
For this we construct a master curve, exploiting the measurements of critical field.
A three-fluid layer model and linear stability analysis on its interfaces allows us to explain the limitation mechanism.
Our results can help in the development of instability based micromixers. 

\end{abstract}



\section{Introduction}
\label{sec:Intro}
Microfluidics has been studied interdisciplinary towards many applications, particularly in life sciences \citep{multidiscipl_medicine}.
It is also interesting from the point of flow behaviour, including rheology \citep{BOOK_GALINDO} and searching for different solutions for micro-mixers \citep{review_mixers2005, review_mixers2011, Chen_mixers}, and the miniaturisation of ﬂuid handling and lab-on-a-chip devices, which are active research topics \citep{CONVERY201976}.

Microfluidics handles liquids within submillimeter ranges, therefore small Reynolds numbers and laminar flows are typical \citep{BOOK_GALINDO}. 
To rapidly enhance otherwise slow and diffusion-limited mixing in a contactless manner, magnetic materials and fields can be used \citep{Chen_mixers}. 


Magnetically actuated particles provide one route to enhancing mixing in laminar microfluidic flows, where molecular diffusion alone can be slow.
A recent review discusses particle based micromixers, including the motion of individual particles under applied fields and the effects of magnetic and hydrodynamic interactions \citep{MisraKumaran2024}.

The magnetic mixing in the scope of microfluidics opens up interesting research topics: various fascinating instabilities emerge on the interface between magnetic and non-magnetic fluids when exposed to an external magnetic magnetic field \citep{Miranda, krakov2021}. 
The nature of the patterns formed by these instabilities depends upon many factors, for example, the direction of the magnetic field, the shape of the interface between the fluids, the size of the microfluidic chip and others \citep{chen_circular, derec_2008, secondary_fingers}. 
Instabilities on an interface between two miscible fluids are a current research topic. 
For example, recently the effectiveness of mixing magnetic and non-magnetic fluids in various magnetic fields was described using numerical simulations, including static field \citep{krakov_2023} and rotating field \citep{KRAKOV2020166186}. 
For nonmagnetic miscible fluids recent examples focus on the viscosity driven finger-like Saffman-Taylor instability in different systems~\citep{NagelPhysRevFluids.4.033902, sharma_nand_pramanik_chen_mishra_2020, zhang2020spontaneous, Primkulov_2026} and are well summarized in a recent review \citep{LuaVermaChen2026}.

The magnetic micro-convection is a finger-like instability that appears on an interface between miscible magnetic and non-magnetic fluids.
In 1980 the fingering was observed for the first time on an immiscible interface \citep{Cebers1980}.
A few years later the fingering on a miscible fluid interface was observed and the topic of magnetic micro-convection was started \citep{cebers_1983}. 
It is caused by a ponderomotive force acting on the magnetic fluid in a homogeneous applied magnetic field.
This force is proportional to the concentration of the magnetic particles in the magnetic fluid and the local gradient of the magnetic field. 
The local gradient of the magnetic field arises from the self-magnetic field of the magnetic liquid. 
The ponderomotive force is potential only when the concentration gradient is collinear to the magnetic field gradient. 
Therefore, if the magnetic field is higher than some critical value, a flow is created by any concentration perturbation that disturbs this collinearity \citep{kitnbergs_JFM_2015}.


It was recently found \citep{NagelPhysRevFluids.4.033902} that, depending on the P\'eclet number, two regimes of viscous fingering in a Hele-Shaw cell are possible, with diffusion controlling the instability at small P\'eclet numbers ($<1000$).
In the problem of magnetic fingering the role of the P\'eclet number is played by the magnetic Rayleigh number, which usually is in the range $[0;1000]$ \citep{Erglis2013612}.
Thus it is expected that magnetic fingering is in the diffusion-dominated regime.


Magnetic fluids have several practical applications \citep{applications}, but magnetic micro-convection itself, in addition to promising applications in fluid mixing, has aroused interest as a fundamental research topic, both experimentally \citep{krakov2021} and theoretically \citep{Miranda2, Miranda, Cuhna}.
In this paper we make a step further, by investigating the interplay of magnetic field and gravity effects, which can stabilise magnetic micro-convection in a vertically placed microfluidic chip.

The development and emergence of the magnetic micro-convection instability is determined by several parameters. 
Many of them have already been explored broadly. 
In 2002 the role of an initially diffused concentration distribution for the magnetic micro-convection in a Hele-Shaw cell was explored numerically by a linear stability analysis \citep{igonin_2002}. 
In the following year, the first high-accuracy direct numerical simulations of the instability on a circular interface was presented \citep{chen_circular}. 
Later, the micro-convective flows were measured for the first time and the significance of gravity for this instability in a Hele-Shaw cell was proposed \citep{Erglis2013612}.
It was discovered that the wavelength of the finger-like instability increases slightly if the thickness of the microfluidic chip, which can be considered as a Hele-Shaw cell, increases \citep{derec_2008}.
This study also explored the change in instability wavelength and the critical magnetic fields.
The theoretical description of magnetic micro-convection in a Hele-Shaw cell has also changed over time.
For example, it was improved by introducing a Brinkman term \citep{kitnbergs_JFM_2015}.
In the same study a larger effective diffusion coefficient was also introduced to the experiments, what was later attributed to a convective motion along the thickness of the cell and can be cancelled if thinner channels are used \citep{KITENBERGS2020166247}.
An alternative way to exclude convective motion, is to turn the cell sideways, keeping the denser fluid on the lower part. 
We approached this experimentally with a continuous microfluidic system in our previous work \citep{Kitenbergs_EPJE}, where a gravity component was also added to the theoretical model, which however remained limited to initially stagnant fluids.
Here we improve the experimental system to achieve initially stagnant fluids.
It is worth mentioning that gravity effects on small scale systems have also been acknowledged by other authors in systems with rollers \citep{chaikin}, filamentary structures \cite{Zhu2026} and falling droplets \citep{KrakovEtAl2025}.

The paper is organized as follows.
Section \ref{sec:model} introduces the theoretical model in a vertical Hele-Shaw cell, describes how spectral method allows to do numerical simulations of the problem and reveals the nonlinear dynamics and the stabilisation of the micro-convection. 
It is followed by section \ref{sec:exp}, which introduces the improved experimental setup and describes experimental observations of magnetic field intensity dependent dynamics of magnetic micro-convection in magnetic fluids with different densities. 
Also, a detailed description of experimental data processing is included. 
A quantitative comparison of results from numerical simulations and experiments is done in section \ref{sec:res}.
This includes defining a mastercurve for the problem, comparing results to other examples in literature, as well as explaining the mechanism of finger growth stabilization with linear stability analysis of a three-layer fluid system  with different magnetic particle concentrations. 
Main conclusions follow in the sect. \ref{sec:concl}, while additional explanatory information for theory and experiments is provided in Appendices~\ref{appA}-\ref{appC}. 
For an easy perception of the experiments and simulations, movies are attached in the Supplementary data.

\section{Model and numerical simulations}\label{sec:model}
\subsection{\label{sec:math_mod} Mathematical model}
\begin{figure}
  \centerline{\includegraphics{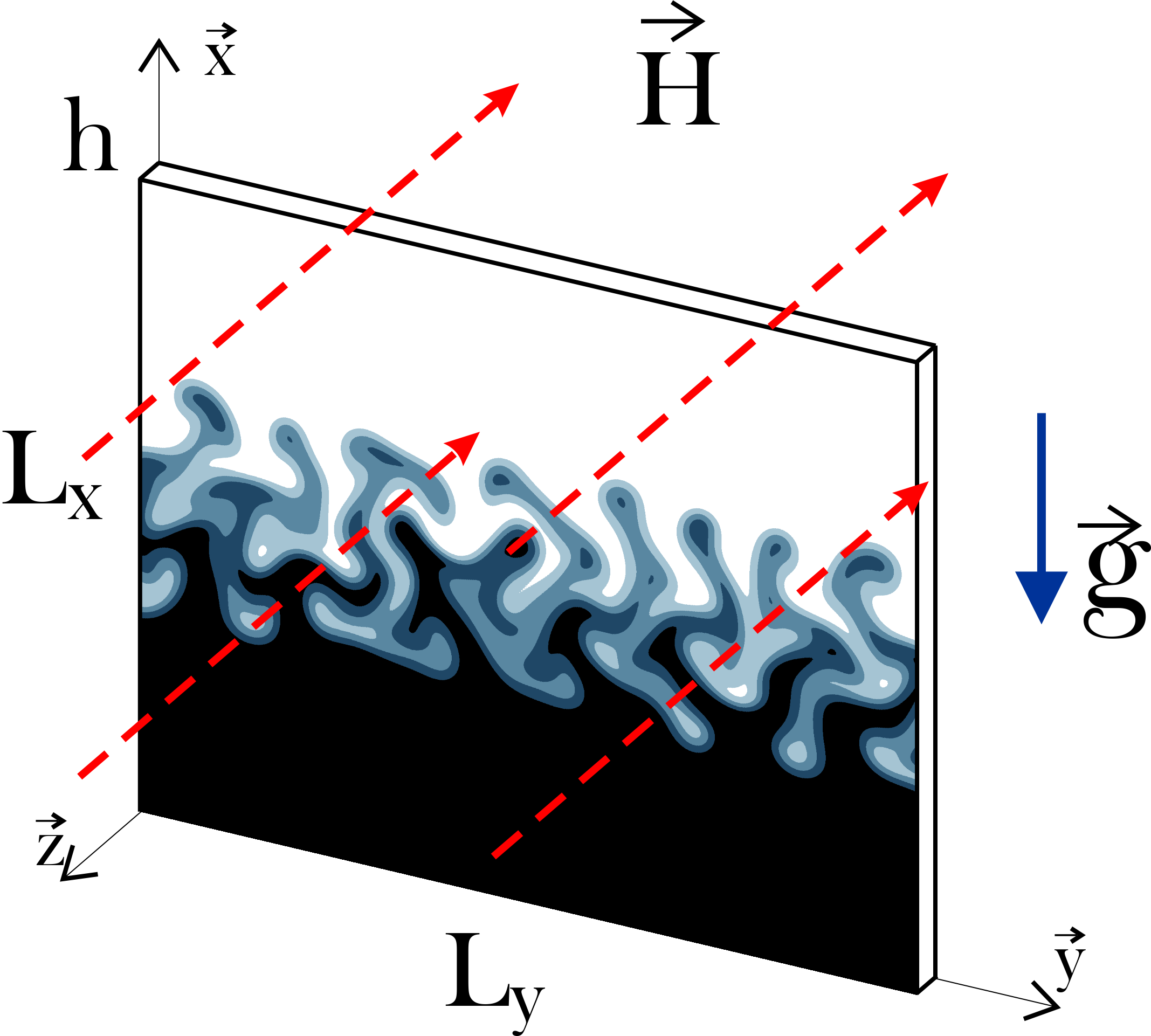}}
  \caption{(Colour online) The sketch of a Hele-Shaw cell for the magnetic micro-convection.}
\label{fig:fig1}
\end{figure}

We consider a model previously presented in \citep{Kitenbergs_EPJE} where two miscible fluids are confined in a vertical Hele-Shaw cell. The magnetic field is applied perpendicularly to the cell, as shown in figure~\ref{fig:fig1}. The upper fluid is nonmagnetic fluid, but the lower one is ferromagnetic fluid.

Due to the ponderomotive forces of the non-homogeneous self-magnetic field on the interface between fluids, a fingering instability arises.

The evolution of the fingering instability is described by a set of equations, which includes the Brinkman equation with magnetic force (\ref{eq:Brinkman}), the continuity (\ref{eq:continuity}) and convection – diffusion equation (\ref{eq:diffusion}) and reads
\begin{equation}
     -\bnabla p -\frac{12 \eta}{h^2} \boldsymbol{v} - \frac{2 M(c)}{h}\bnabla \psi_{\rm m}(c) + \eta{\rm\nabla^2}\boldsymbol{v} + {\rm\Delta}\rho c \boldsymbol{g}  = 0
     \label{eq:Brinkman}
\end{equation}
\begin{equation}
\bnabla \bcdot \boldsymbol{v} = 0
\label{eq:continuity}
\end{equation}
\begin{equation}
\dfrac{\partial  c}{\partial  t} + (\boldsymbol{v} \bcdot \bnabla) c = D \nabla^2 c
\label{eq:diffusion}
\end{equation}
%

Equations (\ref{eq:Brinkman})--(\ref{eq:diffusion}) are two-dimensional: they are written for the depth-averaged velocity in the plane of the cell, so that $\bnabla\equiv(\partial_x,\partial_y)$ and $\nabla^2\equiv\partial_{xx}+\partial_{yy}$.

The viscosities of the two fluids are considered equal.
Here $c$ is the concentration of the magnetic fluid, $\boldsymbol{v}=(v_{x},v_{y})$ is the depth averaged velocity, $\eta$ is the viscosity of the fluid, $p$ is pressure, $h$ is the thickness of the Hele-Shaw cell, $D$ is the isotropic constant diffusion coefficient and ${\rm\Delta}\rho=\rho_{MF}-\rho_{H_2O}$ is the density difference between the fluids.
Magnetisation $M(c)$ is taken to be proportional to the concentration of the magnetic fluid $c$ ($M = M_0 c$) and the value of the magnetostatic potential $\psi_{\rm m}$ on the boundary of the Hele-Shaw cell is given by \citep{Cebers1981}, 
$\psi_{\rm m}(\boldsymbol{r},t) = M_0\int c(\boldsymbol{r}^{'},t) K(\boldsymbol{r}-\boldsymbol{r}^{'},h) {\rm d} S^{'}$
where the integration is performed over the boundary of the Hele-Shaw cell, $ K(\boldsymbol{r},h)=1/\mid \boldsymbol{r}\mid-1/\sqrt{\mid \boldsymbol{r}\mid^2+h^2}$.

The boundary conditions for the velocity components, the concentration of the fluids, and the conditions of the periodicity across the cell require that the fluid is motionless at both ends of the Hele-Shaw cell. The boundary conditions are : $\{c,v_x,v_y\}|_{y=0} = \{c,v_x,v_y\}|_{y=L_y}$ and $\{c,v_x,v_y\}|_{x=0} = \{1,0,0\}$; $\{c,v_x,v_y\}|_{x=L_x} = \{0,0,0\}$, $L_{x}$ and $L_{y}$ are the dimensions of the Hele-Shaw cell figure~\ref{fig:fig1}.

To get the dimensionless form of the Brinkman equation~(\ref{eq:Brinkman}), we use several scales (length $h$, time $h^2/D$, velocity $D/h$ and magnetostatic potential $M_0h$) and introduce the magnetic Rayleigh $\Ram$ and gravitational Rayleigh $\Rag$ numbers: 

\begin{equation}
     -\bnabla p - {\rm \boldsymbol{v}} - 2 \Ram c \bnabla \psi_{\rm m}(c) + \frac{{\rm\nabla^2 }{\boldsymbol{v}}}{12} - \Rag c { \boldsymbol{e}_x}   = 0; \qquad   {\rm\bnabla} \bcdot {\rm \boldsymbol{v}} = 0
     \label{eq:Brinkman:dimen}
\end{equation}

\begin{equation}
\dfrac{\partial c}{\partial t} + (\rm \boldsymbol {v } \bcdot \bnabla) c = \nabla^2 c
\label{eq:diffusion:dimen}
\end{equation}

$\Ram$ is expressed as a ratio between the characteristic time of the diffusion $\tau_{D}=h^2/D$ and the characteristic time of motion that is driven by non-homogeneous self-magnetic
field of the fluid $\tau_{M}=12\eta/M_{0}^2$, and reads as
\begin{equation}
     \Ram=M_{0}^2h^2/12{\eta}D
     \label{eq:Ram}
\end{equation}
$\Rag$ is the ratio between the characteristic time of the diffusion $\tau_{D}$ and the characteristic time of motion due to the gravitational
field $\tau_{G}=12\eta/{\Delta}{\rho}gh$:
\begin{equation}
     \Rag={\Delta}{\rho}gh^3/12{\eta}D
     \label{eq:Rag}
\end{equation}
where ${\Delta}{\rho}=\rho_{MF}-\rho_{H_{2}O}$ is the density difference between the
denser magnetic fluid below and less dense water above and $g$ is the standard gravity.
\subsection{\label{sec:num_sim} The Numerical simulations}
An evolution of the magnetic micro-convection flow in a vertical Hele-Shaw cell is studied numerically in the nonlinear stage using the spectral method in the formulation of the vorticity-stream function~\citep{TanHomsy:PhysFluid:88} \citep{ZimmermanHomsy:PhysRevA:92:46}. When introducing the stream functions $v_x = \displaystyle \frac{\partial \phi}{\partial y}$ and $v_y = -\displaystyle \frac{\partial \phi}{\partial x}$ and the vorticity $\omega = -\nabla^2 \phi$, the convection-diffusion equation~(\ref{eq:diffusion:dimen}) is rewritten using these formulation for the concentration $c(x,y,t) = c_0(x,t) + c^{'}(x,y,t)$.

\begin{eqnarray}
\label{Eg:Num:4}
  \displaystyle \frac{\partial c^{'}}{\partial t} =  \nabla^2 c^{'}   - \mathcal{J}(c,\phi).
 \end{eqnarray} 
 

Here $c_0(x,t)$ is the one-dimensional diffusive base state, which obeys the diffusion equation, and $c'(x,y,t)$ is the concentration perturbation. The initial base profile is $c_0(x,0)=\frac{1}{2}\big[1-\mathrm{erf}\big(x/2\sqrt{t_0}\big)\big]$, where the smearing parameter $t_0$ is the dimensionless diffusion time elapsed before the start of the simulation.

 
The vorticity equation is obtained by calculating the curl of the Brinkman equation (\ref{eq:Brinkman:dimen}) and similarly as for concentration for magnetostatic potential $\psi_m(x,y,t) = \psi_{m0}(x,t) +  \psi_{m}^{'}(x,y,t)$

 \begin{eqnarray}
 \label{Eg:Num:5}
\omega = -\frac{1}{12}\nabla^4\phi - 2 \Ram\mathcal{J}(c,\psi_m) + \Rag \frac{\partial c^{'} }{\partial y} ~,
 \end{eqnarray}
 
 where $\psi_{m0}$,  $\psi_{m}^{'}$  and the Jacobian $\mathcal{J}$ are defined as 

\begin{subeqnarray}
    \psi_{\rm m0}(x,t) & = &\int_{-\infty}^{+\infty} c_0(x-\zeta,t)\ln(1+\zeta^{-2})\mathrm {d} \zeta ,  \\
  \psi'_{\rm m}(\boldsymbol{r},t) & = & \int\limits_{S} c^{'}(\boldsymbol{r}^{'},t)K(\boldsymbol{r}-\boldsymbol{r}^{'},1) \mathrm {d} S^{'}~.
\\
\mathcal{J}(c,f) & = & \displaystyle \frac{\partial(c,f)}{\partial(x,y)} = \frac{\partial c}{\partial x}\frac{\partial f}{\partial y} - \frac{\partial c}{\partial y}\frac{\partial f}{\partial x}, \qquad c=c_0(x,t)+c^{'}(x,y,t)~.
\end{subeqnarray}

The concentration perturbation $c^{'}$, stream function $\phi$, vorticity $\omega$ and the advection and magnetic coupling terms $ J(x,y,t) =   \mathcal{J}(c,\phi)$, $Q(x,y,t) =   \mathcal{J}(c,\psi_m)$ are presented by the Fourier series 
$c^{'}(x,y,t) = \hat{F}[c_{mn}]$, $\phi(x,y,t) = \hat{F}[\phi_{mn}]$, $\omega(x,y,t) = \hat{F}[\omega_{mn}]$, $J(x,y,t) = \hat{F}[J_{mn}]$ and $Q(x,y,t) = \hat{F}[Q_{mn}]$. $\hat{F}$ is Fourier transform operator $\hat{F}[f(x,y,t)] = \sum\limits_{m = 0}^{M-1}\sum\limits_{n = 0}^{N-1} \hat{f}_{mn} (t) \displaystyle{\exp [ \mathrm{i}(q_m x + k_n y)]} $. 

\begin{eqnarray}\label{Eg:Num:Fourier:1}
\hat{\omega}_{mn} = (q_m^2 + k_n^2) \hat{\psi}_{mn}~,
 \end{eqnarray}
 \begin{eqnarray}\label{Eg:Num:Fourier:2}
\hat{\omega}_{mn} = -\frac{(q_m^2 + k_n^2)^2}{12}\hat{\psi}_{mn} - 2 {\Ram}  \hat{Q}_{mn} +  \mathrm{i} k_n {\Rag} \hat{c}_{mn}~,
 \end{eqnarray}
 \begin{eqnarray}\label{Eg:Num:Fourier:3}
\frac{\partial \hat{c}_{mn}}{\partial t} = -(q_m^2 + k_n^2) \hat{c}_{mn} -  \hat{J}_{mn}~.
 \label{Eq:Num:DiffConcen}
 \end{eqnarray}

where $ \displaystyle q_m = \frac{2 \pi m}{L_{x}}$, $ \displaystyle  k_n = \frac{2 \pi n}{L_{y}}$ are wave numbers ($m,n = 0,1,2 \ldots $ ).
The linear differential equation~(\ref{Eq:Num:DiffConcen}) is solved for the known stream function by applying the linear propagator method and discretised by using the three-step Adams-Bashforth method \citep{SamarskijGulin:book}.


The equations are solved pseudospectrally: the nonlinear terms are evaluated in physical space and transformed to Fourier space with the fast Fourier transform, whereas the spatial derivatives in the linear part are computed spectrally.

\begin{figure}
\begin{subfigure}{\textwidth}
\includegraphics[width=\columnwidth]{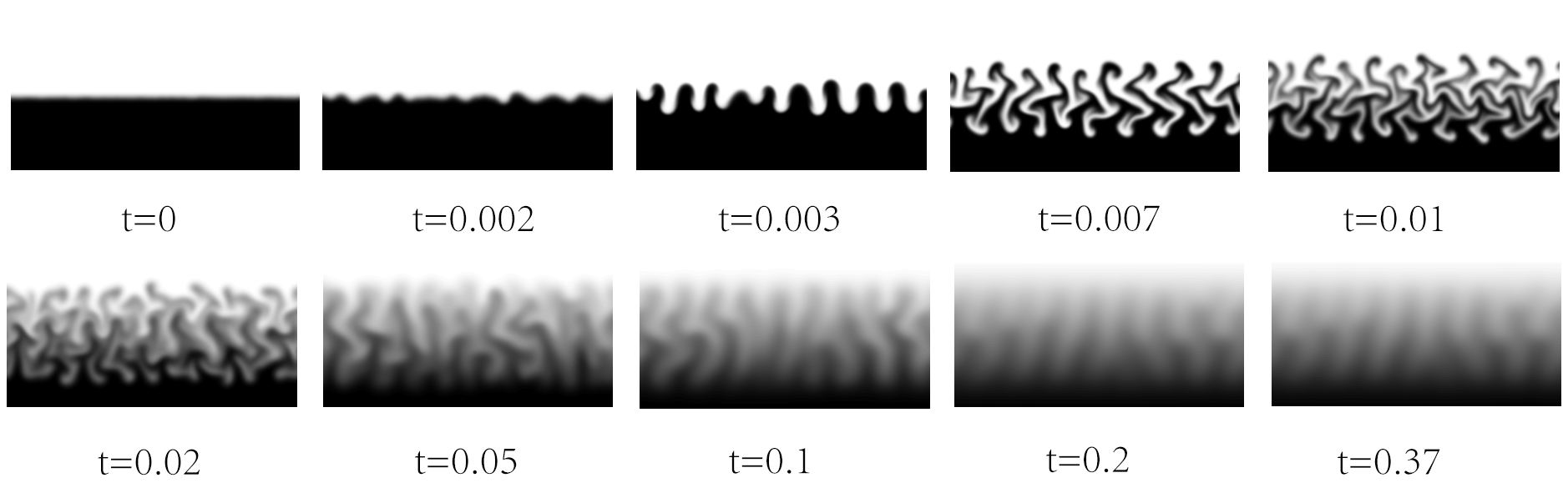}
\caption{$t_0 = 0.0033$}
\label{fig:fig2_a}
\end{subfigure}
\begin{subfigure}{\textwidth}
\includegraphics[width=\columnwidth]{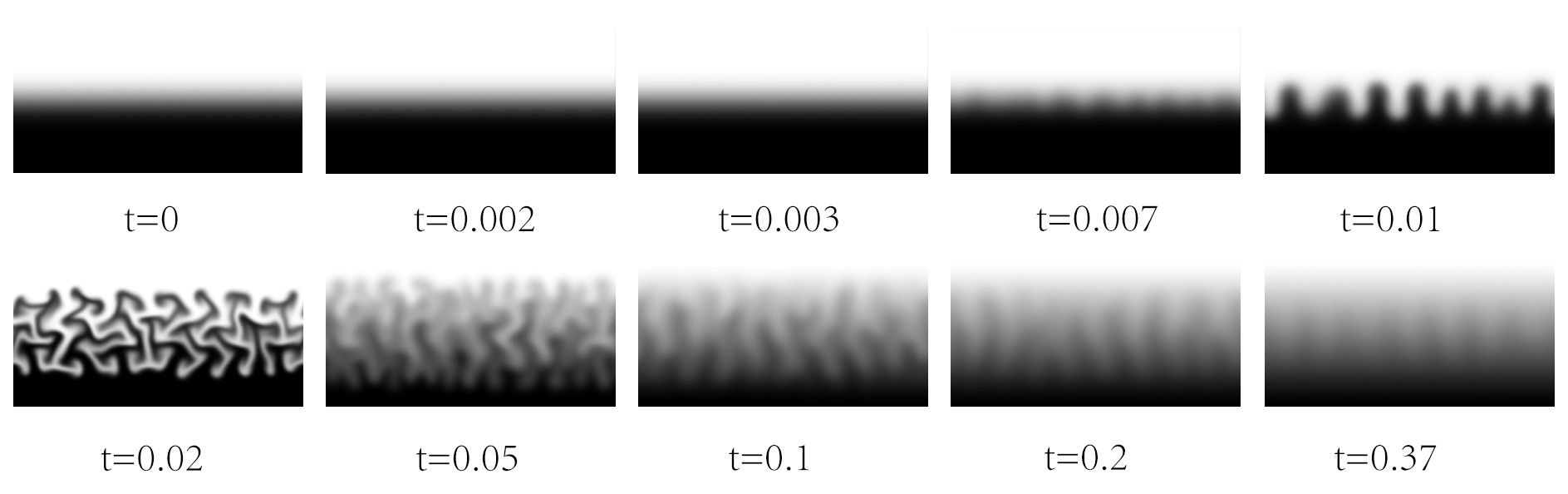}
\caption{$t_0 = 0.1$}
\label{fig:fig2_b}
\end{subfigure}
  \caption{Numerical simulation concentration plots of the magnetic micro-convection dynamics for $Ra_m = 2000$ and $Ra_g = 3031$ at different dimensionless time moments for two values of initial smearing parameter $t_0$.
  Each image has $L_x = 10$, $L_y=20$ size. 
  }
\label{fig:fig2}
\end{figure}

An example of the magnetic micro-convection dynamics from numerical simulations is visible in figure~\ref{fig:fig2}.
Here the concentration plots (water is white, magnetic fluid - black) correspond to $L_x=10$ and $L_y=20$ size.
It can be noted that in each time-step the fingers of the instability are vertically restricted. 
After some time, the fingers stop to growing and only the diffusion remains.

\begin{figure}
  \centerline{\includegraphics[width=\columnwidth]{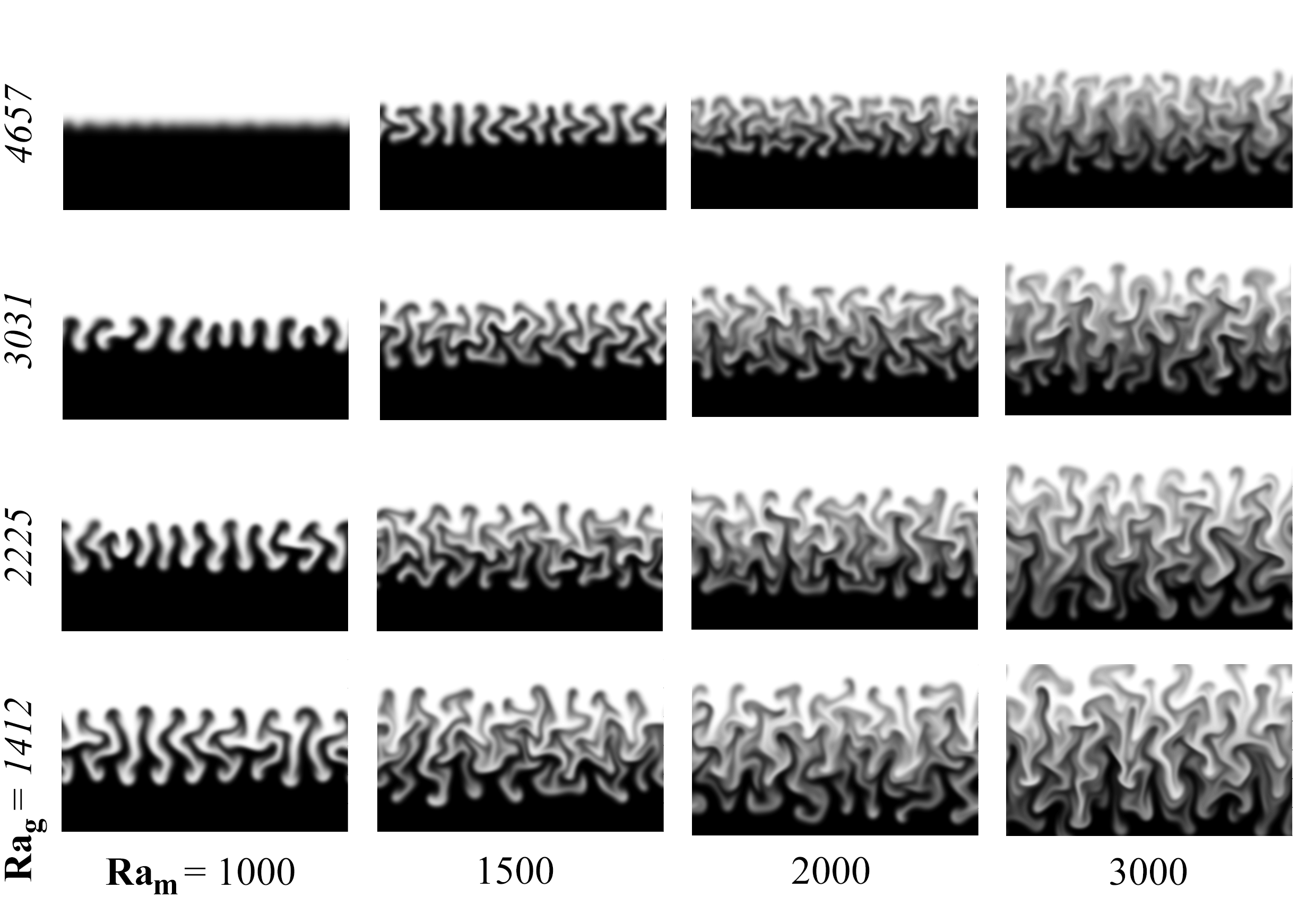}}
  \caption{Numerical simulation concentration plots of the magnetic micro-convection for various $\Rag$ and $\Ram$ values at the same dimensionless time $t=0.015$ and initial smearing $t_0 = 0.0033$. Each image has $L_x = 10$, $L_y=20$ size. 
  }
\label{fig:fig3}
\end{figure}

The way in which the instability is affected by both $\Ram$ and $\Rag$ is compared in figure~\ref{fig:fig3} for a particular dimensionless time $t=0.015$.

The rate of the instability development increases as $\Ram$ increases and is inversely proportional to $\Rag$. 
Gravity effects are observable, as the fingers are less keen to bend and are relatively short for high $\Rag$ numbers. 
Whereas, if $\Ram$ is high and $\Rag$ low enough, even secondary fingers appear.

\section{Experimental system and observations}\label{sec:exp}
\subsection{ Experimental setup}
\begin{figure}
\centerline{\includegraphics[width=0.5\columnwidth]{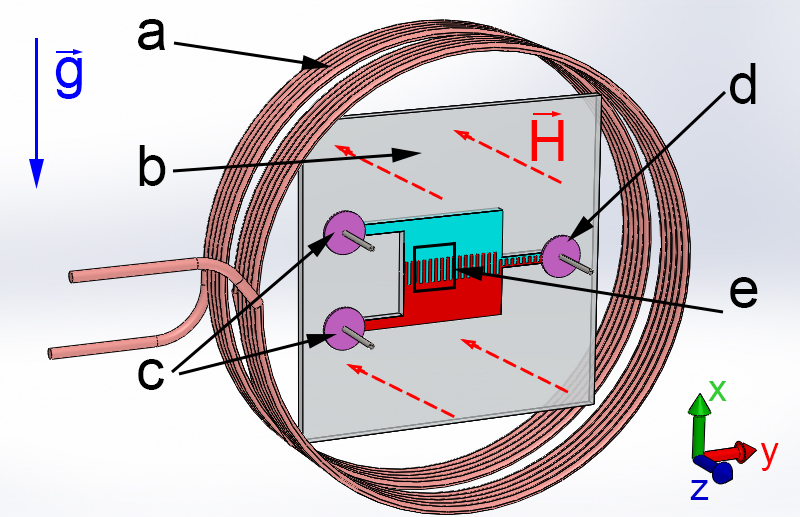}}
\caption{(Colour online) An illustration of the microfluidic chip within a coil system (\textit{a}). A chip (\textit{b}) has two inlets (\textit{c}) and one outlet (\textit{d}) connected to syringe pumps. The lower inlet is for the denser magnetic fluid (red), whereas the upper inlet is for water(blue). The coils provide a homogeneous magnetic field H, perpendicular to the chip. The region of interest (\textit{e}) indicates the area that is analysed.
The microfluidic chip is placed vertically with the direction of gravity field, as shown in the picture. 
}
\label{fig:setup}      
\end{figure}

The setup consists of an optical microscope (Zeiss Stemi 2000-C) with a LED panel (Visional\textregistered, 4~W, 400~lm, 3000~K) as a light source, two syringe pumps (Harvard Apparatus PHD Ultra and KD Scientific Legato 210P), a camera (Lumenera Lu165c, 15~Hz) connected to a computer and a coil system. 
The coil system (see figure~\ref{fig:setup} (\textit{a})) is made from two identical coils and can create a homogeneous magnetic field up to $H = 200$~Oe. 
The coils are powered by a power supply (TENMA 72-2930) in a constant current mode.

The microfluidic chip is vertically fixed in the centre of the electromagnet by a 3D printed holder (PLA filament; Mass Portal Pharaoh XD~20), so that the magnetic field is perpendicular to the plane of the chip. 
The microscope is put sideways and has a connected camera, which records the experiments.

The microfluidic chip (see figure \ref{fig:setup} (\textit{b}) is made of a Paraﬁlm M\textregistered~spacer (thickness $h=0.135\pm0.005$~mm) fixed between two microscope glass slides. 
The microchannel is cut in the spacer with a paper knife, but the inlets and outlets for the fluids are made from syringe tips, which are glued into the holes drilled in one of the glasses. 
This method of producing microchips is fast and inexpensive; however, it is not possible to make two exactly identical microchips. 
A more detailed description of this method can be found in our previous study \citep{Kitenbergs_EPJE}.

There are two innovations in the setup in order to obtain initially stagnant fluids. 
We use two syringe pumps instead of one, and the shape of the microchannel is chosen with a wide rectangular pool. 
The microchannel has two inlets (figure~\ref{fig:setup}(\textit{c}), which are connected to one syringe pump and one outlet (fig.~\ref{fig:setup} (\textit{d})) which is connected to the other syringe pump. 
Pumps are connected electronically with a switch so that they can operate synchronously. 
The pumps start and stop working at the same time, so that the same amount of fluid is pumped into and out of the microchannel. 
This allows one to rapidly stop the flow.

The particular shape of the microfluidic channel is chosen empirically to minimise the flows that exist after pumps have stopped and the magnetic field has been turned on. 
The nature of parasitic flows is not clear. 
Probably, they arise due to the local differences of the demagnetising field.

In experiments we use two miscible fluids — water-based magnetic fluid and distilled water as a nonmagnetic fluid. 
The original magnetic fluid is made by a co-precipitation method~\citep{Massart}, forming maghemite nanoparticles with an average diameter $d=7.0$~nm, volume fraction $\Phi =2.8$~\%, saturation magnetisation $M_{sat}=8.4$~G, magnetic susceptibility $\chi_m^{CGS}$=0.016, determined by a vibrating sample magnetometer (Lake Shore 7404).
Nanoparticles are stabilised with citrate ions.

For liquid handling, we use 1 ml syringes that are connected to the microchip with FEP tubing (IDEX, $\varnothing_{ID} =0.76$~mm, $\varnothing_{OD}=1.59$~mm). 
The magnetic fluid is slightly more dense than water ($\rho_{MF_1}=1.143$~g/cm$^3$), therefore its tubing is connected to the lower inlet of the microchip, while water is connected to the upper inlet.

We perform experiments with 4 different densities of magnetic fluids- noted as $MF_1$, $MF_2$, $MF_3$ and $MF_4$ with densities as follows $\rho_{MF_1}=1.143$~g/cm$^3$, $\rho_{MF_2}=1.091$~g/cm$^3$, $\rho_{MF_3}=1.066$~g/cm$^3$, $\rho_{MF_4}=1.040$~g/cm$^3$. 
To vary the densities, we dilute the original magnetic fluid $MF_1$ with distilled water. 
The density is calculated from a weight measurement with an analytic balance (KERN) for a known volume, taken with a pipette (Gilson). 
When the original magnetic fluid is diluted, its magnetic susceptibility changes proportionally to the magnetic fluid and water volume ratio.

\subsection{Experimental procedure and processing of data}

\begin{figure}
\centerline{\includegraphics[width=.7\columnwidth]{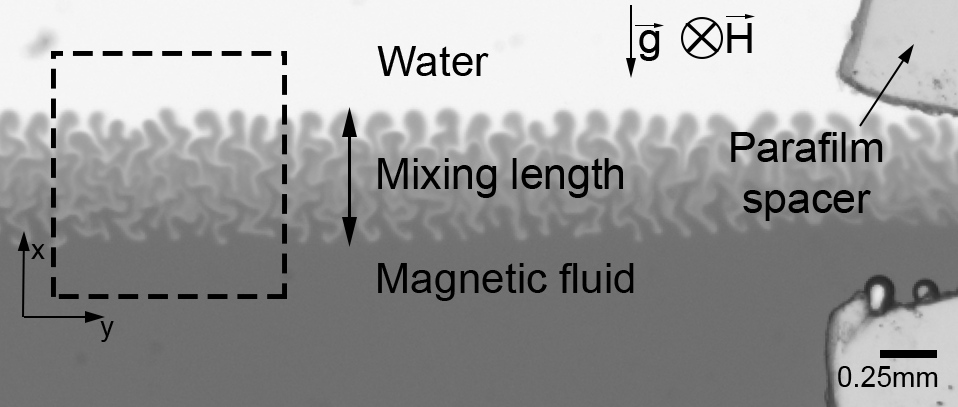}}
\caption{A recorded image from an experiment of the magnetic micro-convection for a magnetic fluid $MF_1$ and water for an external magnetic field $H=89$~Oe at $t=5$~s in a microfluidic chip. The region of interest (area of analysis) is indicated by the dashed line.
}
\label{fig:selectedarea}      
\end{figure}

As mentioned previously, experiments with magnetic fluids with four different densities were carried out. 
First, the pumps run both fluids with a high flow rate for a short time, to obtain a smooth interface between the fluids. 
Then the flow is stopped. 
After a short time, once the interface is stable, the magnetic field is applied. 
The experiment is timed from this moment. 
The development of the mixing of two fluids is recorded as image series ($15$~frames per second) over time.
This is done for various intensities of magnetic field, as well as without an external magnetic field applied for all concentrations.
The recorded image series are analysed for a region of interest, a manually selected area with x- and y-axes as shown in figure~\ref{fig:selectedarea}. 
This area is chosen so that all of the mixing is captured at all times for each experiment. 
To emphasize this phenomenon, we introduce a parameter named mixing length (2$\delta$), which characterises the height of the fingers of the instability. 

\begin{figure}
\centerline{\includegraphics[width=0.7\columnwidth]{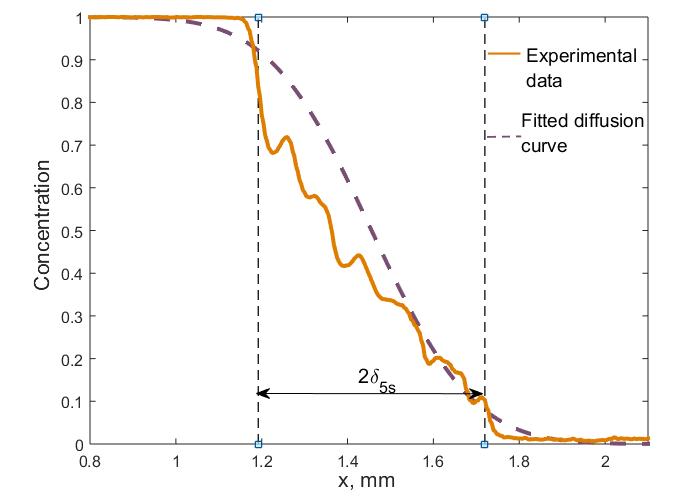}}
\caption{ (Colour online) An average concentration profile (orange line) from experimental data for a time moment $t=5$~s in micro-convection with the magnetic fluid $MF_1$, in external magnetic field $H=89$~Oe and the fitted diffusion curve using Fick's law solution (purple dashed line). The mixing length at $t=5$~s is represented by an arrow as doubled diffusion length $2\delta_{5s}$. }
\label{fig:concentration}      
\end{figure}

To get a quantitative information from the image series, we perform data processing using MatLab.
The steps are described further.
First, each image from the image series is converted from an intensity plot to a magnetic fluid concentration plot c(x,y) via the Lambert-Beer law, which is normalised with respect to initial concentration $c_{0}=1$ \citep{Kitenbergs_JMMM_2015}. 
The concentration of each image is averaged along the y axis, giving $\overline{c}$(x).
X-axis is converted from pixels to mm using a scaling factor determined for the microscope. 
An example of the average concentration profile $\overline{c}(x)$ is shown in figure~\ref{fig:concentration} (see orange line).  

Then for all of the time points this averaged concentration profile $\overline{c}$(x) is fitted with a diffusion curve (see purple dashed line in figure~\ref{fig:concentration}), according to Fick’s law solution:
\begin{equation}
c(x)=\frac{1}{2}(1-\text{erf}(\frac{x-x_{0}}{\delta}))
\label{ficks_law}
\end{equation}
where erf is the error function and $x_{0}$ is the coordinate of the centre of symmetry and gives a degree of freedom for the fit. 
$\delta$ is the diffusion length and equals half of the mixing length (see figure~\ref{fig:selectedarea} and fig.\ref{fig:concentration}). 
For diffusion it is defined as
\begin{equation} 
{\delta}=2\sqrt{Dt},
\label{delta}
\end{equation}
where $D$ is the diffusion coefficient of the magnetic nanoparticles and $t$ is the time during which diffusion is happening.
\begin{figure}
  \centerline{\includegraphics[width=0.6\columnwidth]{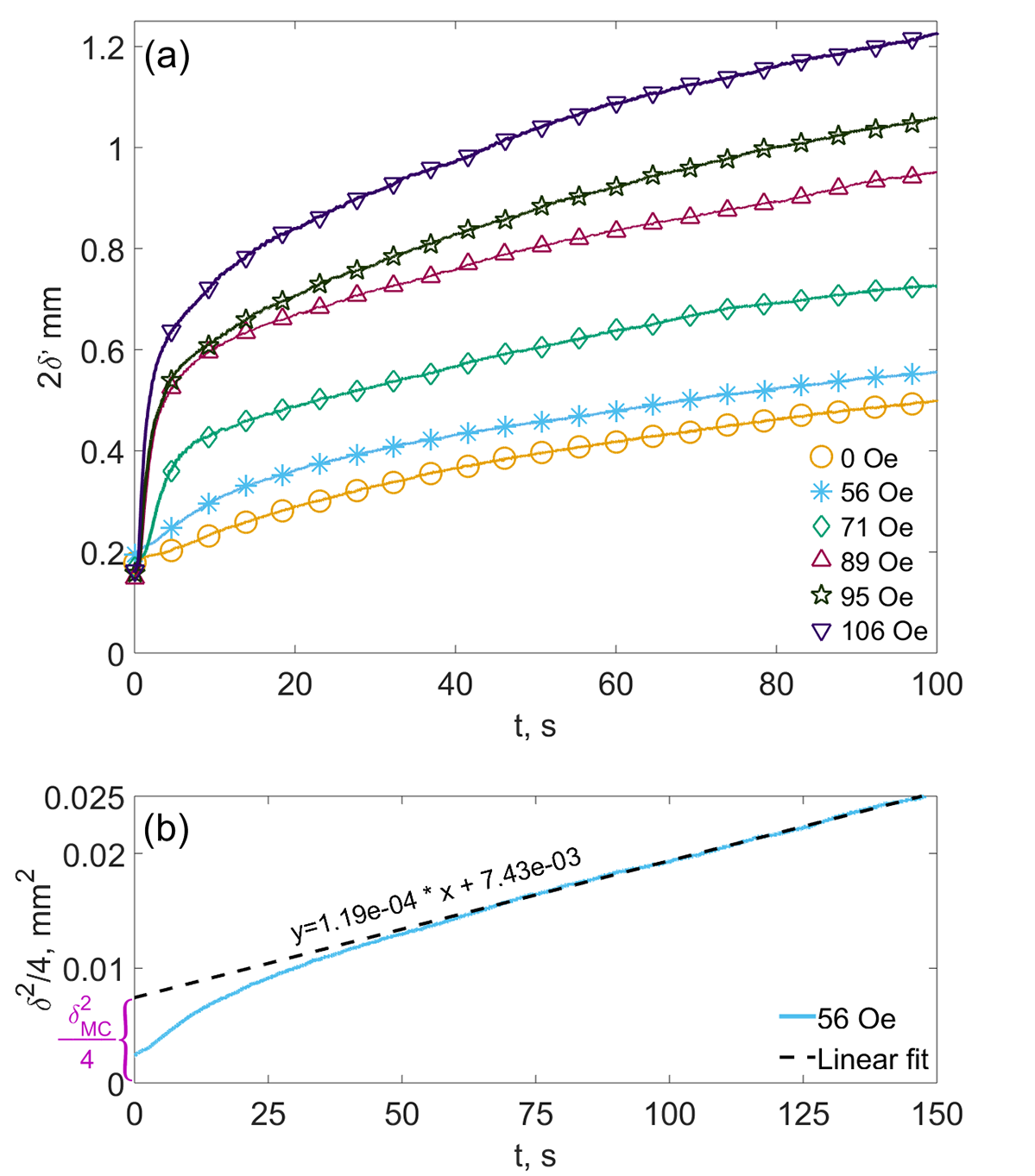}}
  \caption{ (Colour online) The experimentally measured dynamics of fluid mixing. The change in mixing length $2\delta$ over time $t$ for the magnetic fluid $MF_1$ in various magnetic fields is shown in (\textit{a}). $\delta^2/4$ change over time for experiment at $H=56$~Oe is shown in (\textit{b}) as a blue line. The black dashed line is a linear fit from $t=60$~s to $t=150$~s. ${\delta_\text{MC}}^2/4$ is the y-intercept of the linear fit line. The value of the initial interface smearing (value of $\delta^2/4$ at $t=0$~s) is taken into account when calculating the errors. For this experiment ${\delta_\text{MC}}^2/4$~= ${7.4}\cdot{10^{-3}}$~mm$^2$.
  }
\label{fig:delta}
\end{figure}

The next step is to obtain the dynamics of the mixing length over time for a particular experiment. It is done by reprocessing lines of the average concentration profile for all the time frames of each experiment. In figure~\ref{fig:delta} (\textit{a}) a change of mixing length in time can be seen for several experiments.

To identify the diffusive behaviour (using (\ref{delta})), it is useful to plot this dynamic as $\frac{\delta^2}{4}$(t). 
If there is no external magnetic field and the mixing between both fluids is due to diffusion only, then $\frac{\delta^2}{4}$ is directly proportional to time $t$ and the mixing dynamic comes only from diffusion and looks like a straight line.
For the experiments in a magnetic field where the micro-convection is present, this dependency is not linear any more. 
The line has a steep increase at the beginning, but eventually acquires a linear shape. The linearity of the relationship between $\frac{\delta^2}{4}$ and time characterises the diffusion process, but the initial increase is due to the micro-convective instability. 
An example can be seen in figure~\ref{fig:delta} (\textit{b}).

This influence of micro-convection we characterise by a parameter $\delta_\text{MC}$ which we can name micro-convective length. 
The parameter can be obtained by considering the function of the mixing dynamics as a sum of both diffusion and micro-convection. 
Therefore, ${\delta_\text{MC}}^2/4$ is expressed from the difference between the linearly fitted line and the curved line from the experiments. 
Graphically, it is represented by an intercept of the vertical axis at $t=0$~s and a straight line, by which the linear part of the experiment is fitted, as shown in figure~\ref{fig:delta} (\textit{b}).

\begin{figure}
  \centerline{\includegraphics[width=.8\columnwidth]{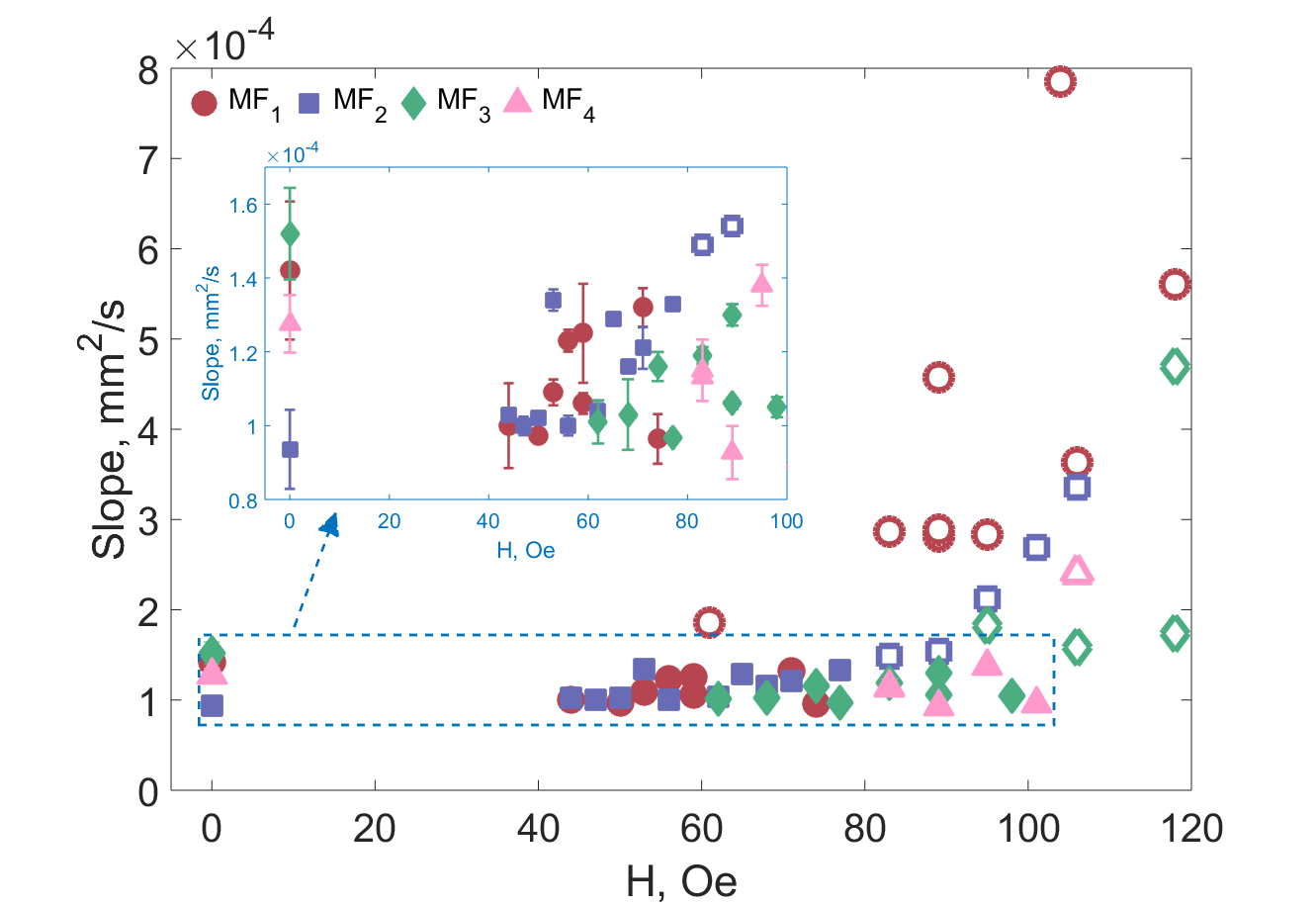}}
  \caption{(Colour online). The slope of the linearly fitted line for experimentally measured $\delta^2/4$ as a function of time for all four magnetic fluids. The empty markers represent that $\delta^2/4$ as a function of time has not reached a linear regime for this experiment. The error bars represent one standard deviation $\sigma$ of the data for the particular experiment.
  }
\label{fig:slope_experimental}
\end{figure}

When $\frac{\delta^2}{4}$ as a function of time is fitted with a line for several experiments at $H=0$~Oe, we find the experimental diffusion coefficient of magnetic nanoparticles $D_\text{exp}=(1.25\pm0.23)\cdot10^{-6}$~cm$^2$/s.
We can use this to verify the quality of our fit.
As can be seen in figure~\ref{fig:slope_experimental}, the experimental values are not the same for higher magnetic fields.
This is because they do not reach the linear regime, and can not be fitted with such a line any more.
Some of the experiments were too short for such a fit, as mixing dynamics did not reach a linear region. 
To estimate the micro-convective length $\delta_\text{MC}$ in such a case, a straight line with a slope $D_\text{exp}$, which comes from experimental data, was attached at the end of the graph.
This way we can assume that the $\delta_\text{MC}^2/4$ should be larger than the y-intercept of the attached line.
For these experiments, the values are represented by the empty markers in figure~\ref{fig:slope_experimental}. 
The analysis of such experiments is described in more detail in the Appendix~\ref{appC}.
Finally, the micro-convective length $\delta_\text{MC}$ for all the experiments is collected and its dependence on the magnetic field and the density of the magnetic fluid is analysed. 
This is described in detail in the results section. 

\subsection{ Experimental observations}

As mentioned previously the timing of an experiment begins when the magnetic field is applied. 
Before that the flow had already stopped for a short time, to stabilise the interface between the fluids. 
This results in an initial smearing up to $0.1$~mm of the interface already before the experiments, as it can be noted in figure~\ref{fig:delta}~a at $t=0$~s. 
This was taken into account when expressing the error bars for the micro-convective length $\delta_\text{MC}$.

\begin{figure}
\centerline{\includegraphics[width=0.9\columnwidth]{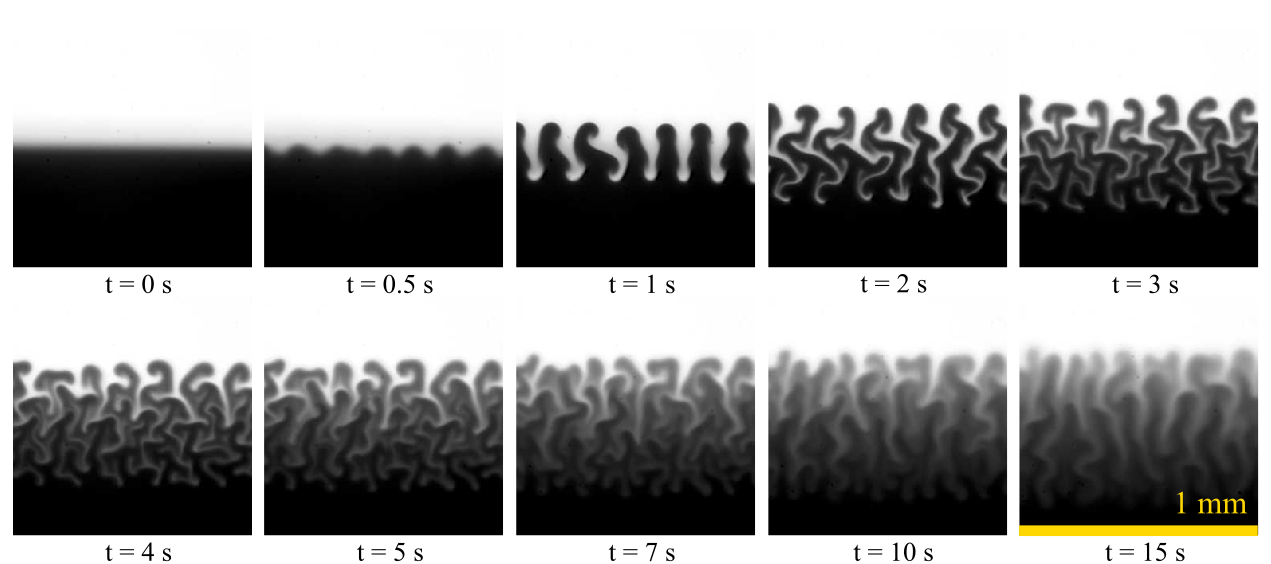}}
\caption{Experimental image series of the magnetic micro-convection dynamics with magnetic fluid $MF_{1}$. The intensity of the applied magnetic field is $H=89$~Oe. The contrast of the images is changed, so that pure water would appear white and pure magnetic fluid black.}
\label{fig:imgrid_exp}      
\end{figure}

When the magnetic field is applied, an instability develops gradually across all fluid interface. 
However, this happens only if the intensity of the magnetic field is above a critical value $H_c$ \citep{Kitenbergs_EPJE}. 
All fingers of the instability grow at the same rate. 
It can be seen in figure~\ref{fig:imgrid_exp}, as at the specific time all fingers have approximately the same height. 
The fingers continue to grow until they reach some maximum height. 
If the intensity of the magnetic field is relatively small, the fingers stay straight and do not branch, not even after some time, as it can be seen in figure~\ref{fig:imgrid_all} ($H_c=41.4$~Oe) and in the movie~1 ($H=47$~Oe) in the supplementary data.
At a slightly stronger magnetic field, fingers already grow considerably taller and start to bend after some time, or already start to grow bent (figure~\ref{fig:imgrid_all}: $H$=$53.2$~Oe and $H$=$61.3$~Oe and movie~2 for $H=50$~Oe).
The fingers branch if the magnetic field is even stronger (figure~\ref{fig:imgrid_all}: $82.8$~Oe and $118.3$~Oe and movie~3 for $H=89$~Oe).
The fingers appear earlier and grow more rapidly in experiments with a higher intensity of the applied magnetic field. 
It can be clearly seen in the movie~3 ($MF_1$ at $H=89$~Oe) where the fingers appear notably earlier ($t=0.3$~s) than at smaller magnetic fields, whereas, for example, at $H=47$~Oe this time is approximately $t=4$~s and at $H=50$~Oe it already is $t=2$~s for $MF_1$. 
It can be clearly seen in movie~3 that after some time ($t=20$~s) the fingers have reached their maximum height, although the convective motion is still present and the fingers continue to bend and branch. 
This is common for experiments at high magnetic fields.
Once the finger has grown, its shape remains the same. 
The edges of the fingers gradually blur due to diffusion. 
This happens until only a smeared area remains. 

Other factors that affect the dynamics of the magnetic micro-convection are the density and the magnetic susceptibility of the magnetic fluid. 
For example, the fingers continue to mix for considerably longer periods of time, if both the intensity of the magnetic field and the susceptibility (in our case related to the density) of the magnetic fluid are higher. 
Comparison of experiments at the same magnetic field $H=106$~Oe, but for different magnetic fluids can be seen in movie~4 ($MF_{1}$) and movie~5 ($MF_{2}$). 
Both movies seem similar, but the fingers grow more rapidly for the denser fluid, which also has a higher susceptibility. 
At the end of the movies, the fingers have reached their maximal height of the micro-convection. 
Nevertheless, the mixing length continues to increase as the diffusion is still present. 
For the diluted fluids, the intensity of the magnetic field must be higher, in order to observe the micro-convection. 
As a result, a higher current must be used and the coils system heats up faster. 
To avoid system overheating, shorter experiments are performed for diluted fluids.

Experiments where notable parasitic flow appears are not analysed. 
Sometimes it is difficult to determine whether parasitic flow appeared at the end of the experiment.
A way to differ if it happens, is from the form of the fingers, which bend specifically in the flow. 
However, when both fluids are considerably mixed, there are no apparent fingers left for clues. 

\begin{figure}
  \centerline{\includegraphics[width=0.8\columnwidth]{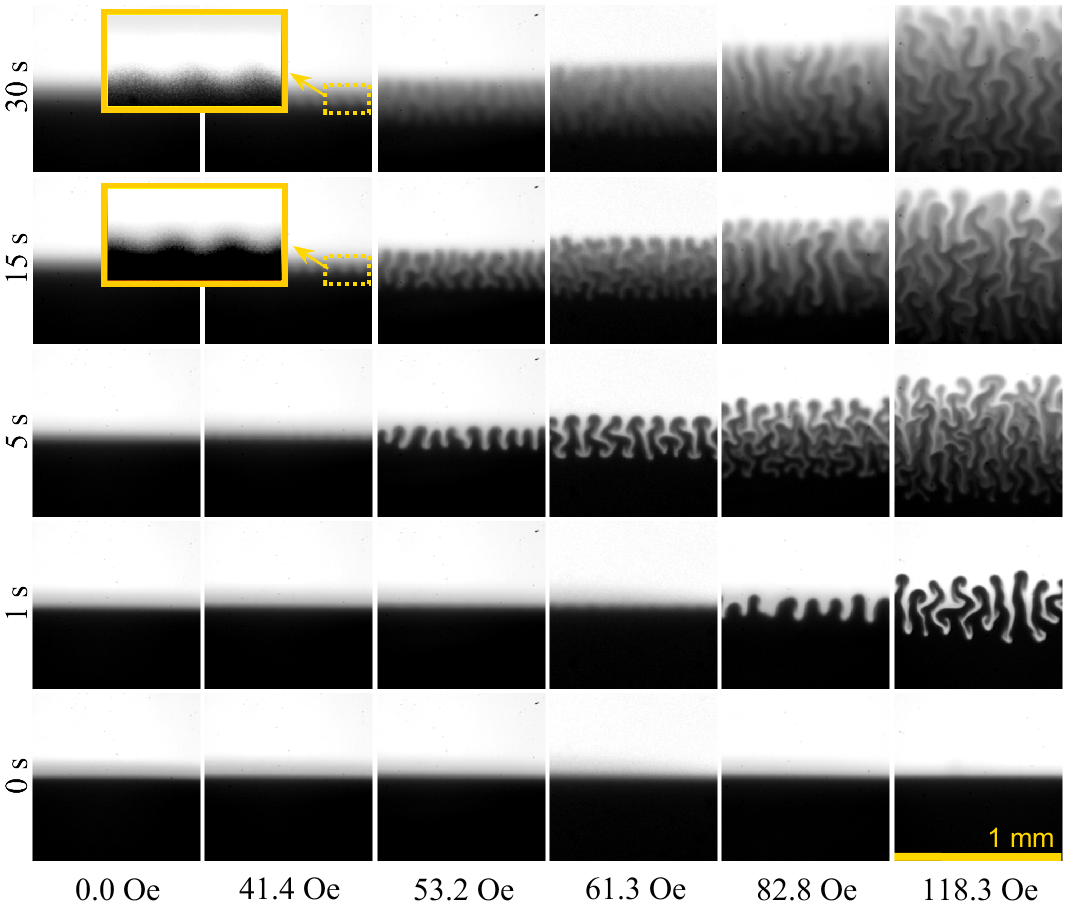}}
\caption{ Experimental images of the magnetic micro-convection dynamics with magnetic fluid $MF_1$ in various magnetic fields. The contrast of the images is changed, so that pure water would appear white and pure magnetic fluid black.
}
\label{fig:imgrid_all}      
\end{figure}
%

\section{\label{sec:res}Results and discussion}
The nature of the micro-convection qualitatively agrees between experiments and numerical simulations, as can be observed in figures \ref{fig:fig2} and \ref{fig:imgrid_exp}, as well as the movies in the Supplementary data.
Movies $1$ till $5$ demonstrate the dynamics of micro-convection experimentally (on the left side of the movie) and in the numerical simulations (on the right side of the movie). 
In all fingers appear after applying magnetic field, and they grow until they reach some maximal height. 
After emerging, fingers have uniform height, and they tend to bend and branch if the magnetic field is strong enough, until the mixing has reached a stage when they remain stationary and only diffusion takes place.
This can be seen both in experiments and simulations. 
There are some visual differences in movies as the experimental movies are recorded as intensity plots, but the movies for the numerical simulations show the concentration plots.

\begin{figure}
  \centerline{\includegraphics[width=\columnwidth]{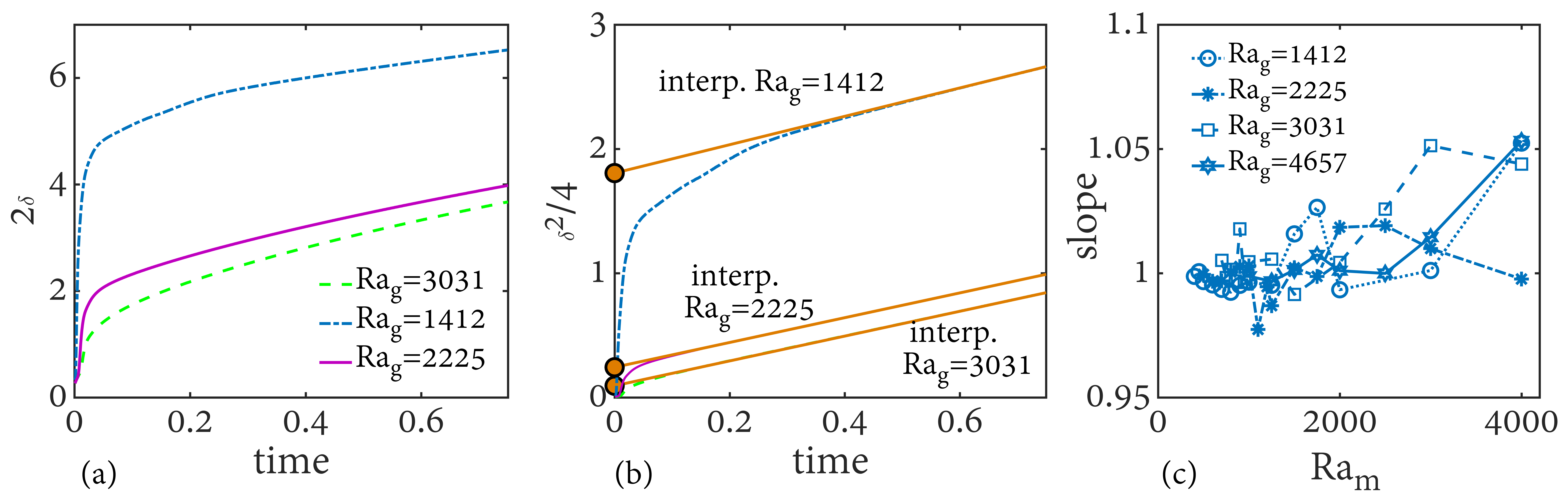}}
  \caption{ (Colour online) Examples of numerical simulation results show the dimensionless mixing length $2\delta$ dependence from dimensionless time $t$ (\textit{a}) and the $\delta^2/4$ dependence from dimensionless time $t$ with interpolation to obtain dimensionless micro-convective lengths at y-axis intercept (\textit{b}) for several $\Rag$ values at $\Ram=800$. The resulting slope (dimensionless diffusion coefficient $D_{sim}$) 
 (\textit{c}) dependence on magnetic Rayleigh number $\Ram$ for different gravitational Rayleigh numbers: ($\Rag=1412$; $\Rag=2225$; $\Rag=3031$; $\Rag=4657$)
  }
\label{fig:fig11}
\end{figure}

To compare the results quantitatively, we find the micro-convective length $\delta_\text{MC}$ also for the numerical simulation data. 
For an example, the dynamics of the magnetic micro-convection from numerical simulations is demonstrated in figure~\ref{fig:fig11}~(\textit{a}) as the change of the mixing length $2\delta$ over time $t$. 
By graphing this relationship as a function of time $\delta^2/4$, a dimensionless diffusion coefficient $D_{sim}$ and a dimensionless micro-convective length $\delta_\text{MC}$ can be obtained in the same way as in the experiments (see figure~\ref{fig:fig11} (\textit{b})). 
Data of the diffusion coefficient is collected in figure~\ref{fig:fig11}~(\textit{c}), showing a good agreement with the expected value~$1$.

An alternative approach to characterize dynamics is to extract growth increments, however this is currently limited to the initial linear process \citep{KlevsEtAl2024}.

\begin{figure}
  \centerline{\includegraphics[width=.8\columnwidth]{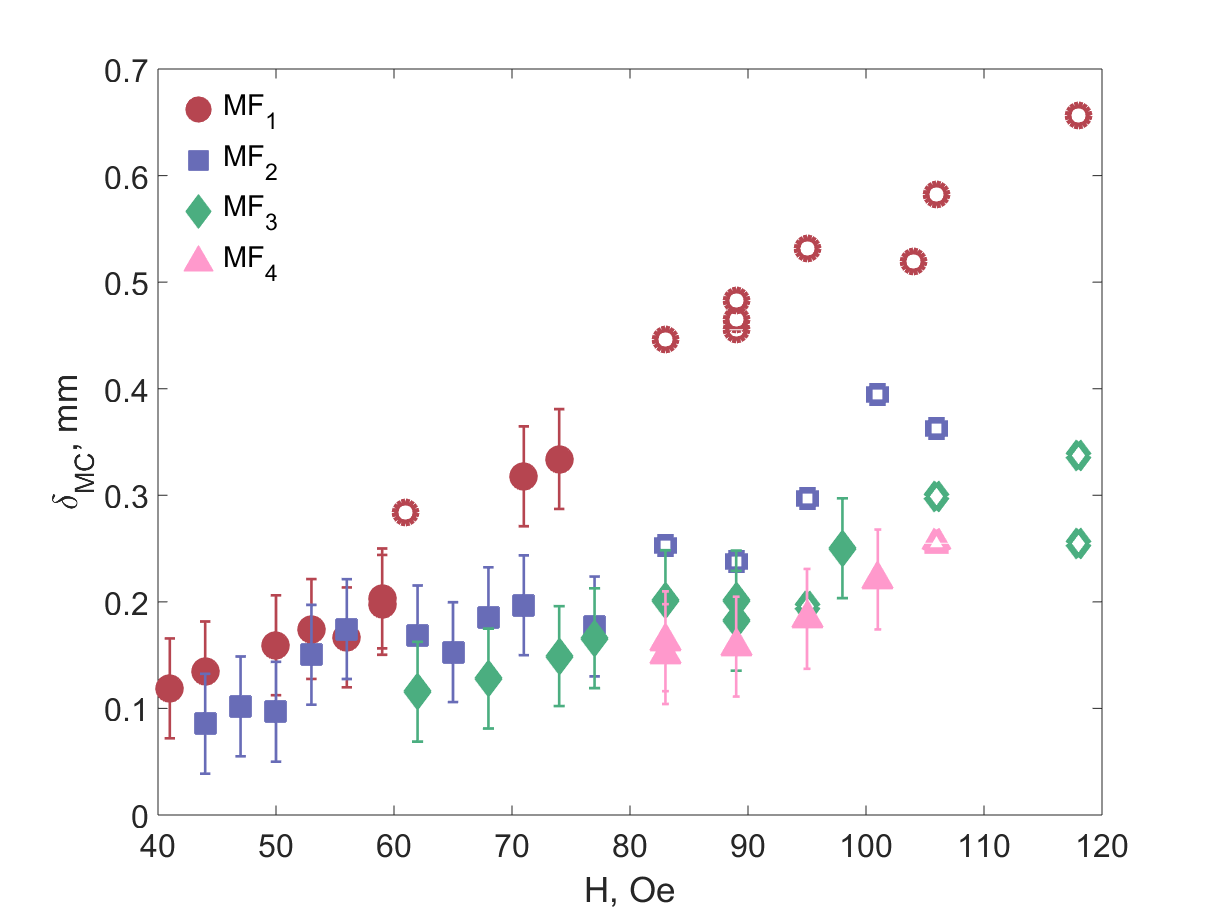}}
\caption{(Colour online) Change of experimentally measured $\delta_\text{MC}$ due to the intensity of the magnetic field H for all four magnetic fluids: $\rho_{MF_1}=1.143$~g/cm$^3$, $\rho_{MF_2}=1.091$~g/cm$^3$, $\rho_{MF_3}=1.066$~g/cm$^3$, $\rho_{MF_4}=1.040$~g/cm$^3$. The empty markers represent that $\delta_\text{MC}$ should be bigger than this value. The error-bars characterise the resolution of the experimental setup for these experiments.}
\label{fig:deltaMC_oe}      
\end{figure}

A quantitative look on the clear field dependence of experimentally observed magnetic micro-convection can be seen in figure~\ref{fig:deltaMC_oe}.
It shows the magnetic micro-convection length $\delta_\text{MC}$ dependence on the intensity of the magnetic field $H$ for different magnetic fluids.
The filled markers represent data obtained from the graphs where the linear part of $\delta^2/4$ as a function of time corresponds to $D_\text{exp}$ within the error limits, while the empty markers correspond to the cases where linear regime has not been reached (See Appendix \ref{appC} for more details). 
One can see that $\delta_\text{MC}$ increases with the intensity of the magnetic field. 
Also, at the same $H$, micro-convective length $\delta_\text{MC}$ is smaller for more dilute magnetic fluids. 
The errors of the experiments are mostly created by the initial interface smearing.

To compare experimental and theoretical results quantitatively, experimental values are converted to dimensionless quantities. 
The experimental density of the magnetic fluids relate to the gravitational Rayleigh number $\Rag$, while the intensity of the external magnetic field is expressed by the magnetic Rayleigh number $\Ram$. 
Though it is interesting to note that $\Ram$ is expressed through magnetisation, which correlates to the concentration of the magnetic fluid and susceptibility in turn. 
It means that in our study, where we use the same magnetic fluid and dilute it with water, to obtain magnetic fluids with various densities, at the same magnetic field $H$ the magnetic Rayleigh number $\Ram$ is higher for the most dense fluid.

For both $\Rag$ and $\Ram$ the channel thickness $h=0.0135$~cm, viscosity $\eta=0.01$~P and the diffusion coefficient $D=5.7\cdot10^{-7}$~cm$^2$/s are used. 
Such a value of $D$ is selected as it complies well with various experimentally obtained values of the diffusion coefficient for the particular magnetic fluid \citep{KitenbergsThesis} or $D_{theor}$, estimated from the Stokes-Einstein equation:
\begin{equation}
     D=\frac{kT}{3{\pi}{\eta}d}
     \label{eq:Stokes}
\end{equation} 
where $k$ is the Boltzmann constant, $T=293$~$K$ the fluid temperature (room temperature) , $\eta=0.01$~P is the viscosity of the water and $d=7.0$~nm is the average diameter of the magnetic nanoparticles. 
This coefficient is two times smaller than the experimentally estimated $D_\text{exp}$. 
The difference might come from a small parasitic convective movement of fluids in the experiments, although visually they appear to be stagnant. Data of the dimensionless diffusion coefficient $D_{sim}$ from numerical simulations and experimental diffusion coefficient $D_\text{exp}$ are collected in figures \ref{fig:slope_experimental} and \ref{fig:fig11} (\textit{c}). 
The value of $D_{sim}$ is 1 in dimensionless units. 

The corresponding values of the gravitational Rayleigh number $Ra_g$ to the experimentally used magnetic fluid $MF_1$ is $\Rag_{1}=4657$, for $MF_2$ is $\Rag_{2}=3031$, for $MF_3$ is $\Rag_{3}=2225$ and for $MF_4$ is $\Rag_{4}=1412$.

\begin{figure}
 \centerline{\includegraphics[width=0.8\columnwidth]{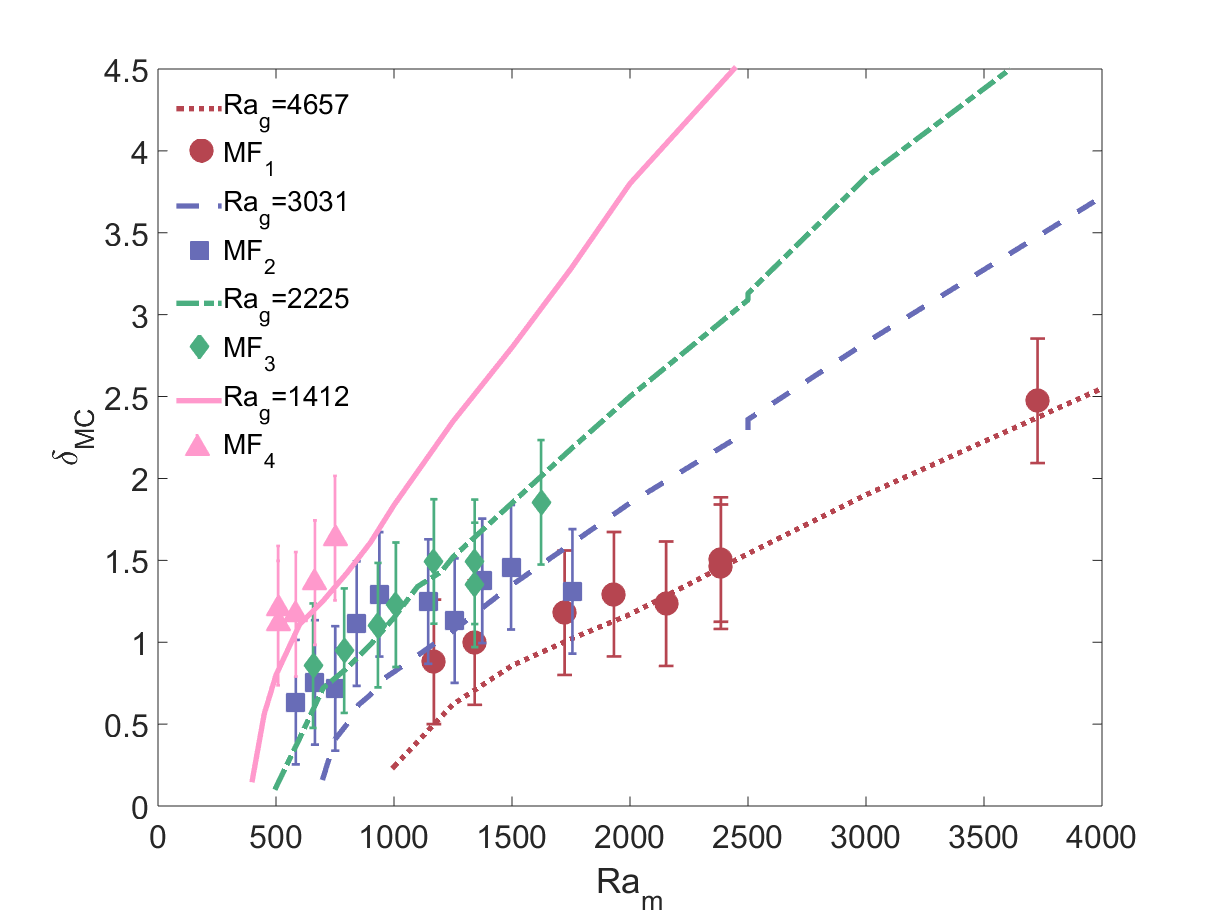}}
\caption{(Colour online) Comparison for experimental data with numerical simulation for $\delta_\text{MC}$ as a function of $Ra_m$. The numerical simulations are shown as lines with the same colour as the corresponding experiment. The error-bars characterise the precision of the experimental setup, as well as initial smearing of the interface.}
\label{fig:comparison_sim_exp}      
\end{figure}

The comparison of mixing enhancement by magnetic micro-convection of experimental and numerical data is shown in figure~\ref{fig:comparison_sim_exp} as $\delta_\text{MC}$ dependence of $\Ram$ in dimensionless units. 
The nature of the numerical and experimental results qualitatively seem the same. 
Gravity plays a higher role for denser fluids (higher $\Rag$), restricting $\delta_\text{MC}$. 
The results of numerical simulations fit well with experimental results with magnetic fluids $MF_1$, $MF_3$ and $MF_4$ within experimental errors and are close also for $MF_2$. 
Due to low contrast and large initial smearing of the interface, it is harder to collect experimental data at small $\Ram$. 
The average initial smearing for these experiments is $\overline{\delta_{t=0s}}=0.08\pm 0.02$~mm, where the error is one standard deviation.

The initial smearing might be considered as a diffusion that occurs for some time. In this way, we calculate the time at $\delta_{t=0s}$ using (\ref{eq:diffusion}) and we find $t_0=29\pm14$~s, which is $t_{0}=0.10\pm0.05$ as dimensionless quantity. 
The effect of the magnitude of the initial smearing to the dynamic of the magnetic instability is shown in movie~6 (magnetic fluid $MF_1$, $H=50$~Oe, $\Rag=4657$, $\Ram=1705$) in the supplementary data. 
The instability is simulated with different initial smearing values (from $t_0=0.075$ till $t_0=0.135$) and the experimental movie with $\delta_0=0.094$~mm (which corresponds to $t_0=0.073$) is added in the lower right corner for comparison in this movie. 
As can be seen in this movie, the instability emerges earlier if the premixed layer of both fluids is thinner.

Experimentally determined characteristic wavelength of the fingers in the beginning of the instability is $\lambda_c=0.12\pm0.02$~mm and does not seem to be dependent on $Ra_g$. 
The $\lambda_c$ corresponds to a dimensionless wave-number $k=7.3\pm1.0$.
We can note that $\lambda_c$ is in good agreement with results in our previous study \citep{Kitenbergs_EPJE} where we experimentally measured $\lambda_c=0.15\pm0.05$~mm for the most concentrated magnetic fluid ($Ra_g=4657$) and water interface. 
Here we should note that in \citet{Kitenbergs_EPJE} the instability started continuously from an initial contact point of the two fluids since both fluids were pumped with the same flow rate continuously. 
In comparison, for experiments in initially stagnant fluids reported here, the instability develops at once across a long interface, which is the same as in the theoretical model in the linear analysis.
\begin{figure}
 \centerline{\includegraphics[width=0.8\columnwidth]{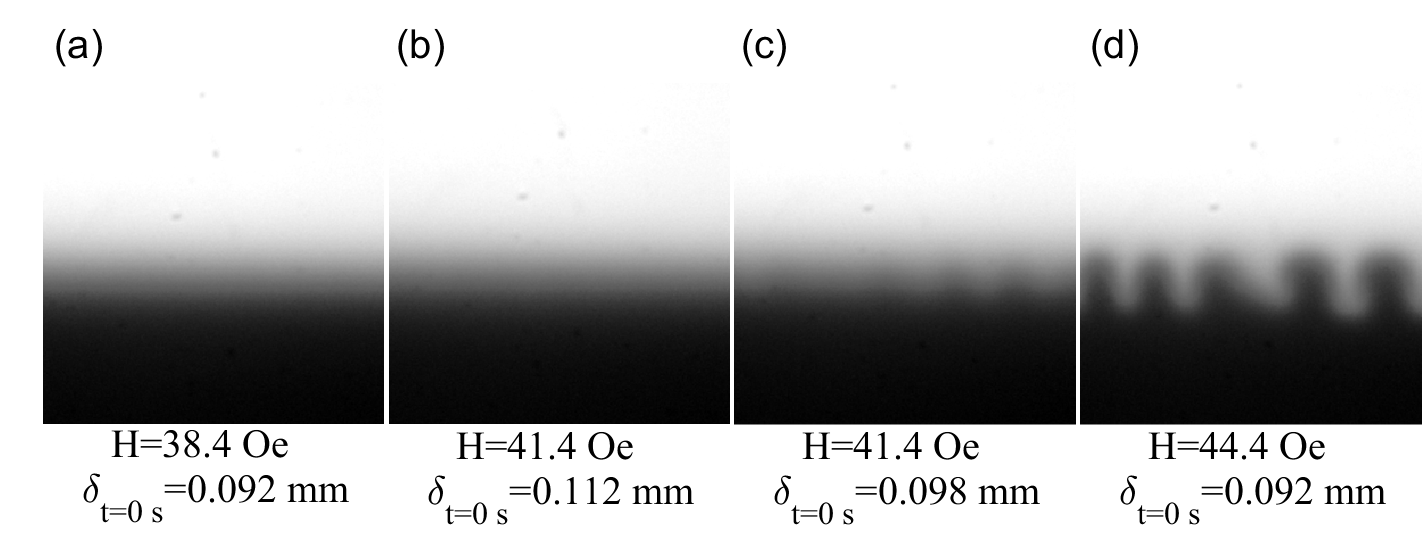}}
\caption{The evaluation of the experimentally measured critical magnetic field $H_c$ for magnetic fluid $MF_{1}$. Pictures were taken at $t=10$~s. The critical magnetic field is estimated to be $H_c=41.4$~Oe (c). The fingers can not be seen in the (b), although $H=H_c$, because the initial smearing $\delta_{t=0~s}=0.112$~mm is larger than in (c), where $\delta_{t=0~s}=0.098$~mm.}
\label{fig:critical_field}      
\end{figure}

Another parameter that affects the precision of the comparison between the experiment and the numerical simulation is the thickness of the microchip $h$. 
Both $\Ram$ and the dimensionless time parameter change when converting experimental units to dimensional units if the thickness $h$ is altered. 
As the time parameter affects also the initial smearing $t_0$ parameter and the rate of the instability within the simulation changes. 

Within this study we also evaluated the critical magnetic field $H_c$ and compared it with the previous estimation \citep{Kitenbergs_EPJE} and the numerical simulations ($H_c$ converted to $\Ram_c$ for comparison). The values obtained for magnetic fluids used are summarised in table \ref{table:critical_fields}. 
As can be seen in figure~\ref{fig:critical_field}, the estimation of $H_c$ experimentally is not unambiguous.
In particular, it is affected by the initial smearing as the fingers appear at higher magnetic fields if the initial smearing is larger. 
Unfortunately, it is impossible to carry out an experiment without an initial smearing with the current experimental system. 
Therefore, we included the critical $\Ram$ values for the instability to emerge from numerical simulations for various initial smearing values $t_0$. 
If the initial smearing is larger, the critical value of $\Ram$ increases. 
Experimentally, the upper value of $H_c$ is visually estimated from experiments, where it is the lowest magnetic field at which the fingers are visible. 
However, fingers might appear even at smaller magnetic fields if the interface between both fluids was sharper at the beginning of the experiment.

\begin{table}
  \begin{center}
\def~{\hphantom{0}}
\begin{tabular}{lcccc}
Source                             & \begin{tabular}[c]{@{}c@{}}$MF_1$\\ $Ra_g=4657$\end{tabular}               & \begin{tabular}[c]{@{}c@{}}$MF_2$\\ $Ra_g=3031$\end{tabular}             & \begin{tabular}[c]{@{}c@{}}$MF_3$\\ $Ra_g=2225$\end{tabular}             & \begin{tabular}[c]{@{}c@{}}$MF_4$\\ $Ra_g=1412$\end{tabular}             \\[3pt]

Numerical simulation, $t_0=0.0033$ & 687  & 487  & 364  & 255 \\ 
Numerical simulation, $t_0=0.001$  & 662 & 477 & 360  & 250   \\
Numerical simulation, $t_0=0.01$   & 725   & 499    & 382  & 263  \\ 
Numerical simulation, $t_0=0.1$    & 1085   & 733   & 554  & 370 \\ 
 \begin{tabular}[c]{@{}c@{}} Linear stability analysis \\ \citep{Kitenbergs_EPJE} \end{tabular}    & 655   & 458   & 355   & 246  \\ 
\begin{tabular}[c]{@{}c@{}} Estimated experimentally \\ \citep{Kitenbergs_EPJE} \end{tabular}         & \begin{tabular}[c]{@{}c@{}}775\\ ($35\pm2$~Oe)\end{tabular}              & \begin{tabular}[c]{@{}c@{}}473\\ ($41\pm1$~Oe)\end{tabular}            & \begin{tabular}[c]{@{}c@{}}395\\ ($50\pm2$~Oe)\end{tabular}            & \begin{tabular}[c]{@{}c@{}}328\\ ($69\pm2$~Oe)\end{tabular}            \\ 
  Experiments, this study        & \begin{tabular}[c]{@{}c@{}}$1170\pm170$\\ ($41.4\pm1.5$~Oe)\end{tabular} & \begin{tabular}[c]{@{}c@{}}$598\pm85$\\ ($44.4\pm1.5$~Oe)\end{tabular} & \begin{tabular}[c]{@{}c@{}}$596\pm80$\\ ($59.1\pm1.5$~Oe)\end{tabular} & \begin{tabular}[c]{@{}c@{}}$473\pm55$\\ ($79.8\pm1.5$~Oe)\end{tabular} \\ 
\end{tabular}
\caption{The critical values of $Ra_m$ (and $H_c$ for experiments) for the micro-convection to emerge.}
\label{table:critical_fields}
\end{center}
\end{table}

From data in table~\ref{table:critical_fields} we can note several things.
First, the values from experiments in this study are slightly larger than estimated in \citet{Kitenbergs_EPJE}.
Difference comes from larger initial smearing in experiments here, while earlier estimations came from sharp interfaces in continuous flow microfluidics.
This is further confirmed with numerical simulation results for various $t_0$ values.
We see that estimations from \citet{Kitenbergs_EPJE} have critical $\Ram$ values closer to small $t_0$ cases, while experimental results in this study are larger and closer to the largest numerical simulation $t_0=0.1$ example. 
Furthermore, the average initial smearing $\overline{\delta_{t=0s}}=0.08$~mm experimentally converts to the previously mentioned $t_0=0.1$ in the simulations.
The influence of the initial smearing on the critical field will be investigated more broadly in a separate publication. 
We can also see that the critical values of $\Ram$ and $H_c$ are influenced by gravity - they increase with an increasing $\Rag$.

\begin{figure}
 \centerline{\includegraphics[width=0.8\columnwidth]{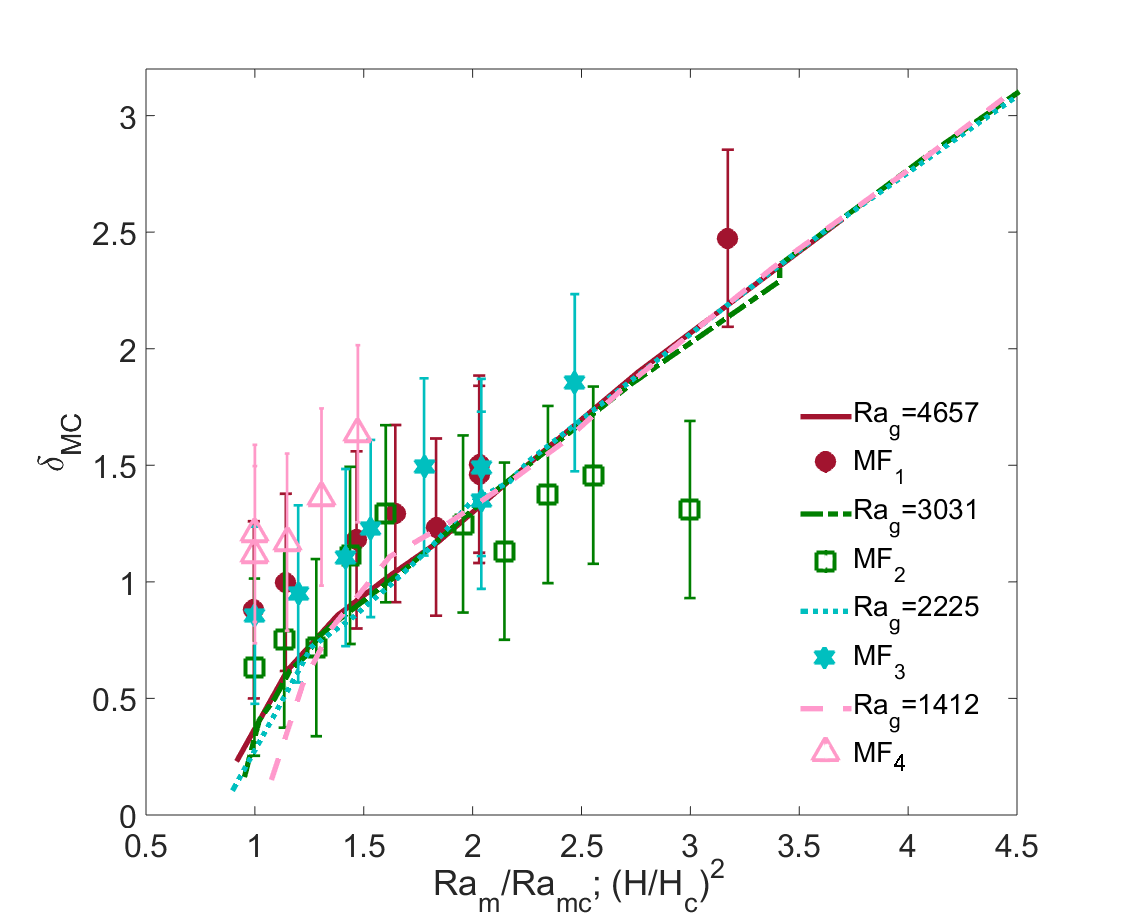}}
\caption{(Colour online) Defining a master curve for the development of non-dimensional $\delta_\text{MC}$. The numerical results are described with $Ra_m/Ra_{mc}$, but the experiments with $(H/H_c)^2$ - both of these axis coincide. The numerical simulations are shown as lines, with the same colour as the corresponding experimental markers.}
\label{fig:mastercurve}      
\end{figure}

An interesting approach to analysing the micro-convective length $\delta_\text{MC}$ field dependence is by dividing $\Ram$ with the critical value $\Ram_c$, which we can see in figure~\ref{fig:mastercurve}. 
This proportion make numerical data for all $\Rag$ coincide. 
Even more, knowing that $\Ram$ is proportional to $H_c^2$, if for experimental measurements the magnetic field is divided with the experimental critical field $H_c$ and this division is squared, while normalizing dimensional $\delta_\text{MC}$ with thickness of the cell $h$, also the experimental results arrange themselves around the same curve - we have found the mastercurve. 
The shape of the curve has a sharper increase for $1<\Ram/ \Ram_c<1.5$ and becomes close to linear for $\Ram / 
 \Ram_c>1.5$. 
Our hypothesis is that the critical values of the magnetic field (or $\Ram$ in numerical simulation case) include significant information about the gravity effects of the micro-convection development, which in practice allows to cancel the gravity effect on $\delta_\text{MC}$ dynamics. 

\subsection{Theoretical model for mixing limit}

\begin{figure}
 \centerline{\includegraphics[width=0.8\columnwidth]{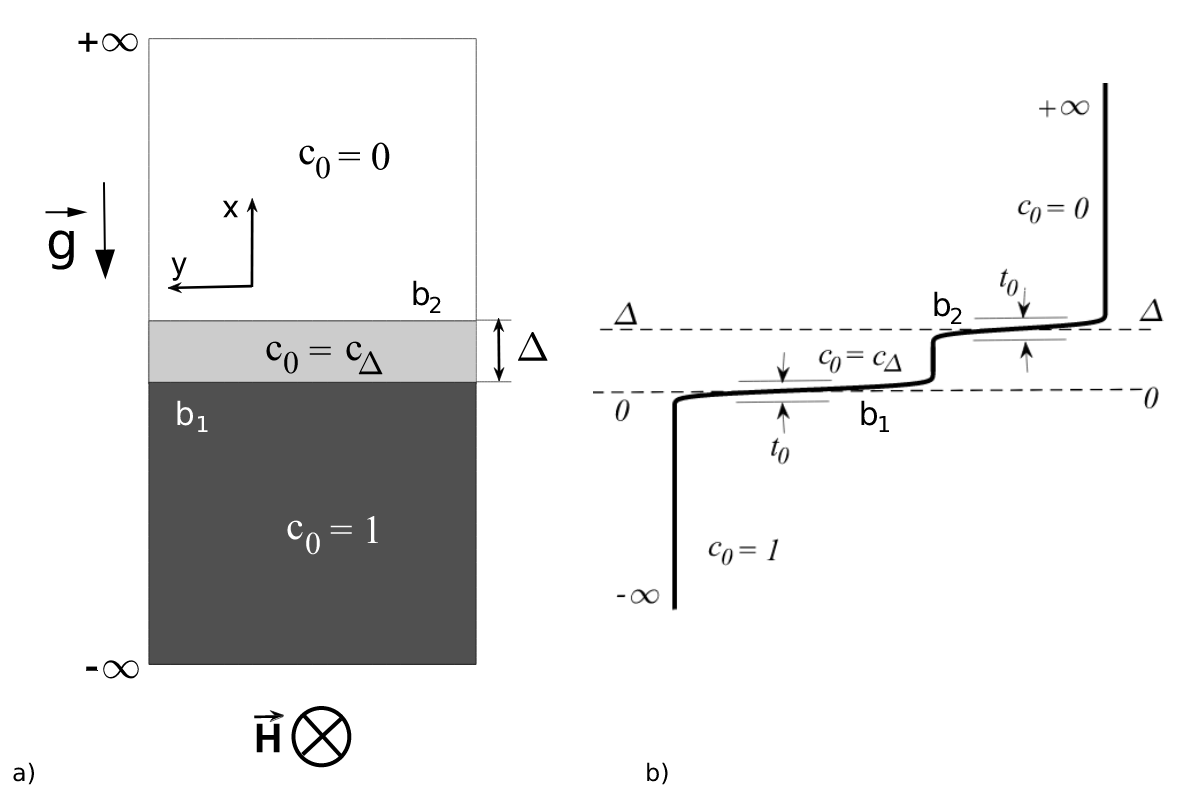}}
\caption{A schematic illustration of the Hele-Shaw cell with three layers of miscible magnetic fluids - a). The initial concentration of the magnetic fluid for the corresponding layer is constant $c_0 = \left[1,c_\Delta,0\right]$, while $b_1$ and $b_2$ mark respective boundaries - b). The external magnetic field $\mathbf{H}$ is normal to the Hele-Shaw cell.}
\label{fig:LinAnaliz:1}       
\end{figure}

To look for a theoretical explanation of the characteristic vertical limit at which the micro-convective mixing stops, we use a very simplified approach. 
We define a system with three layers as shown in figure~\ref{fig:LinAnaliz:1}, where a layer of concentration $c_0=c_\Delta$ and thickness $\Delta$ is located between the initial fluids with concentrations $c_0=1$ and $c_0=0$. 
We hypothesize that finding the vertical limit of convective mixing is a similar problem as to find the critical field for a mixed interface with some larger thickness. 
Therefore, we solve the stability problem of the middle layer, shown in fig.~\ref{fig:LinAnaliz:1}, in the case of Brinkman flow Eq. (\ref{eq:Brinkman:dimen}) and Eq.(\ref{eq:diffusion:dimen}) . 

An analytical solution exists in the limit for the smearing parameter on the boundaries between liquids when $t_0 \rightarrow 0$, that is, when the concentration distribution on the boundary between the two fluids is step-like. The linear perturbation of a quiescent base state is represented by:

$\{u_x,u_y,c,\psi_{\rm m} \}(x,y,t) = \{0,0,c_0,\psi_{\rm m0} \}(x) + \{u_x^{'},u_y^{'}  ,c^{'},\psi_{\rm m}^{'} \}(x)\exp[\displaystyle \mathrm{i} k y + \lambda t]$ 
where magneto-static potential 
$\psi_{\rm m0}(x,t_0) = \int_{-\infty}^{+\infty}c_0(x+\xi,t_0)\ln(1+\xi^{-2}){\rm d}\xi
$  and initial distribution of concentration $ c_0 = 0.5\left(1- (1 - c_{\Delta}){\rm erf}\left( \displaystyle {x}/{(2\sqrt{t_0})}\right) - c_{\Delta}{\rm erf}\left( \displaystyle {(x-\Delta)}/{(2\sqrt{t_0})}\right)\right) $.

By adding the linear perturbation of the velocity, the concentration and the magnetic potential, into Brinkman, continuity and concentration equations, and by neglecting the second and any higher-order terms we get:

\begin{eqnarray}
\label{Eq:Lin:Pert1}
 (\lambda + k^2 )c^{'} + u_x^{'} \frac{\partial c_0}{\partial x} - \frac{\partial^2 c^{'}}{\partial x^2}  = 0 ~.
\end{eqnarray}

\begin{eqnarray}
\label{Eq:Lin:Pert2}
\left( \frac{\partial^2}{\partial x^2} - k^2 \right )^2 u^{'}_x - 12 \left( \frac{\partial^2}{\partial x^2} -  k^2 \right )  u^{'}_x  - 24  k^2 \textit{Ra}_{ m}   \left( \frac{\partial c_0}{\partial x}\psi_{\rm m}^{'} -   c^{'}\frac{\partial \psi_{\rm m0}}{\partial x} \right) +  12 k^2 \textit{Bg} c^{'} = 0
\end{eqnarray}

For step-like $c_0(x)$ the boundary conditions are obtained by integrating the equations (\ref{Eq:Lin:Pert1}) and (\ref{Eq:Lin:Pert2}) across the diffusion layer $\int_{-\delta}^{+\delta}[...] dx$ and taking the limit $\delta \rightarrow 0$. The details of the solution are described in the appendix~\ref{appA}. 

\begin{figure}
 \centerline{\includegraphics[width=0.7\columnwidth]{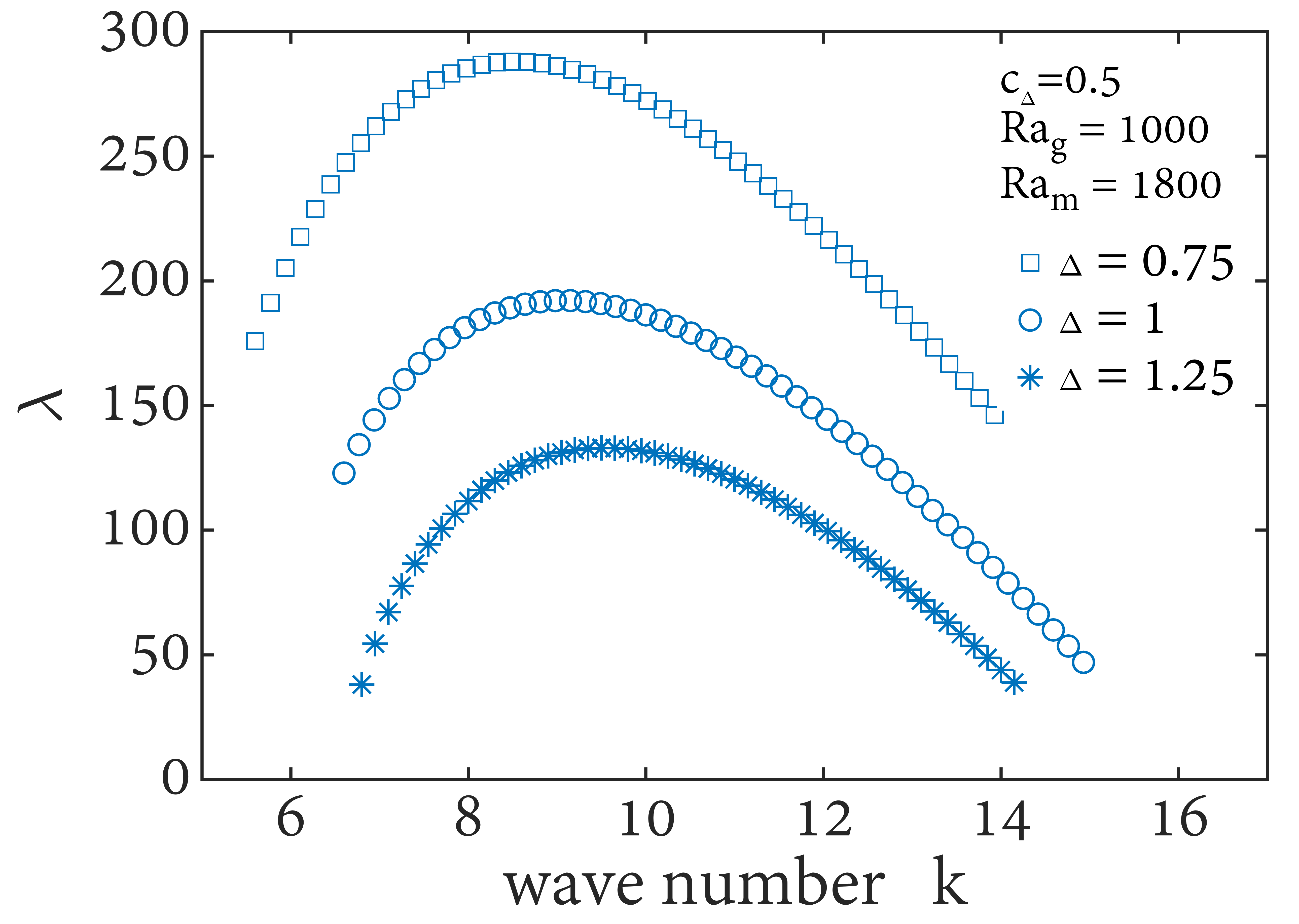}}
\caption{(Colour online) Growth increments of the normal field instability at the interface in middle layer of three miscible fluids in dependence on the wavenumber at different values of layer width $\Delta = 0.5; 1; 1.25$. $\Rag = 1000$, $\Ram = 1800$ and $c_{\Delta} = 0.5$. 
}       
\label{fig:LinAnaliz:FastestWaveNumber}
\end{figure}

\begin{figure}
 \centerline{\includegraphics[width=.6\columnwidth]{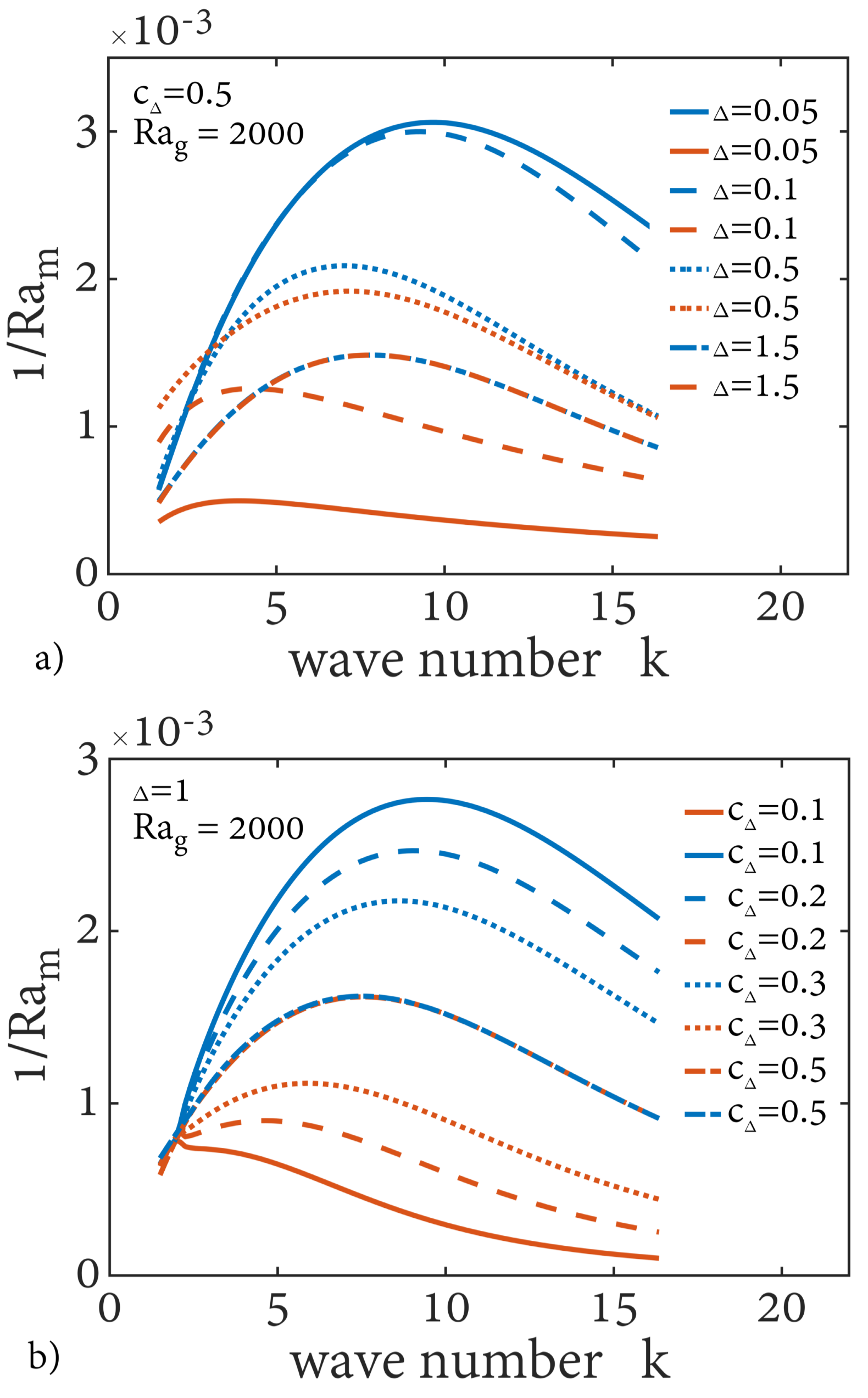}}
\caption{(Colour online) The neutral stability curves ($\lambda = 0$) of the magnetic micro-convection for two boundaries and a mixed layer. a) Four different values of layer width $\Delta = 0.05; 0.1; 0.5; 1.5 $ at $Ra_g = 2000$ and $c_{\Delta} = 0.5$. \\b) For different values of magnetic particle concentrations in the mixed layer $c_{\Delta} = 0.1;0.2;0.3;0.5$ at $Ra_g = 2000$ and $\Delta = 1$.\\Blue lines indicate values for boundary $b_1$ (magnetic fluid - mixed layer), orange lines - boundary $b_2$ (mixed layer - water)}
\label{fig:LinAnaliz:Neutral:1}        
\end{figure}

\begin{figure}
 \centerline{\includegraphics[width=0.7\columnwidth]{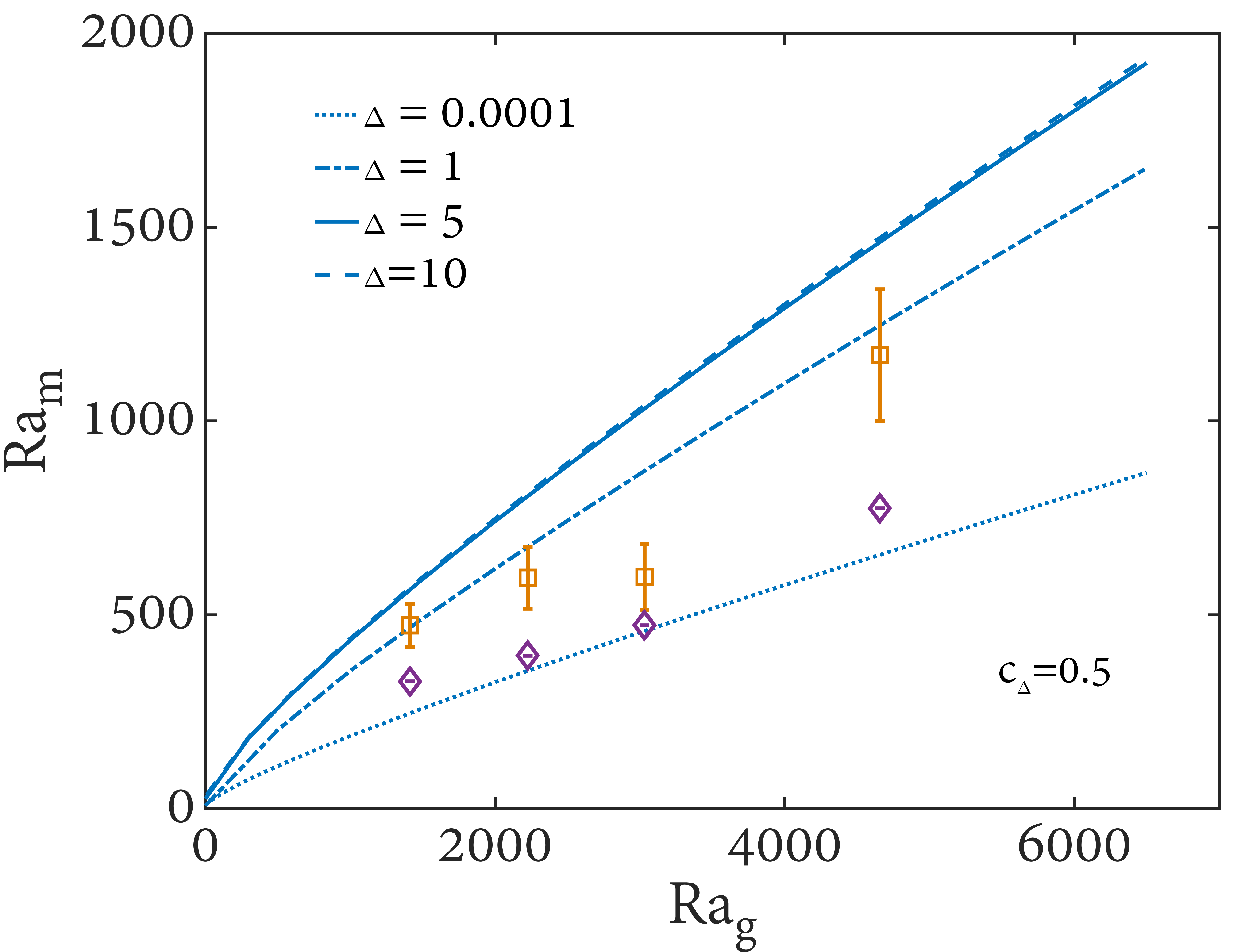}}
\caption{(Colour online) The critical magnetic Rayleigh number $\Ram$ dependence of the gravitational Rayleigh number $\Rag$ at different values of  layer thickness  $\Delta = 0.0001; 1; 5; 10$ (lines) ,   concentration of magnetic particle in middle layer is $c_{\Delta} = 0.5$ and experimental data for two values of initial smearing $t_0 = 0.01$ (romb, from \citep{Kitenbergs_EPJE}) and $t_0 = 0.1$ (square, this study).
}       
\label{fig:LinAnaliz:CritRamRagDelta}
\end{figure}

The linear analysis was carried out for various thicknesses $\Delta$ and concentrations of the middle liquid layer $c_\Delta$.
The growth increment of small perturbations as a function of the wave number $k$ for different values of thickness $\Delta = 0.5; 1; 1.25$ and for the concentrations of magnetic particles in the middle layer is $c_{\Delta} = 0.5$  (the gravitational and magnetic Rayleigh numbers  are $\Rag = 1000$ and $\Ram = 1800$)  is shown in figure~\ref{fig:LinAnaliz:FastestWaveNumber}. 
Thus we see that the increase of the layer thickness stabilizes the instability.
The maximum values of growth increment for observed  magnetic and  gravitational Rayleigh numbers lie in the interval between $k \simeq 8$ and $k \simeq 10$, what is close to experimental observation. 
The results of the neutral curves of instability can be seen in figure~\ref{fig:LinAnaliz:Neutral:1}.
Maximum of each curve gives the critical Rayleigh number $\Ram_c$ for figure~\ref{fig:LinAnaliz:CritRamRagDelta} for the particular parameters and interface. 
One can notice that critical values agree on both interfaces $b_1$ (boundary between the magnetic fluids) and $b_2$ (boundary between the magnetic fluid and non-magnetic fluid) if the thickness of the layer is large enough ($\Delta>1$) and $c_\Delta=0.5$.

Using this approach, we find the critical magnetic Rayleigh number $\Ram$ dependence from gravitational Rayleigh number $\Rag$ for a variety of thicknesses of middle layer $\Delta$ and plot them in fig.~\ref{fig:LinAnaliz:CritRamRagDelta}.
We can see that for a given value of magnetic field increase in gravity force will reduce the thickness of the layer which is stable.
With increasing the thickness of layer the instability of the boundary  between phases becomes independent of distance from second boundary due to decreasing magnetic interaction between the magnetic particles and can be observed simply as for two phase  model.
We can compare this with the experimental data for critical magnetic Rayleigh numbers for two values of initial smearing parameter $t_0 = 0.1$ (this study) and $t_0 = 0.01$  (data from \citet{Kitenbergs_EPJE}).
This is shown in figure~\ref{fig:LinAnaliz:CritRamRagDelta} with squares and rombs respectively and is close to the results from the linear stability analysis.    

\begin{figure}
 \centerline{\includegraphics[width=0.99\columnwidth]{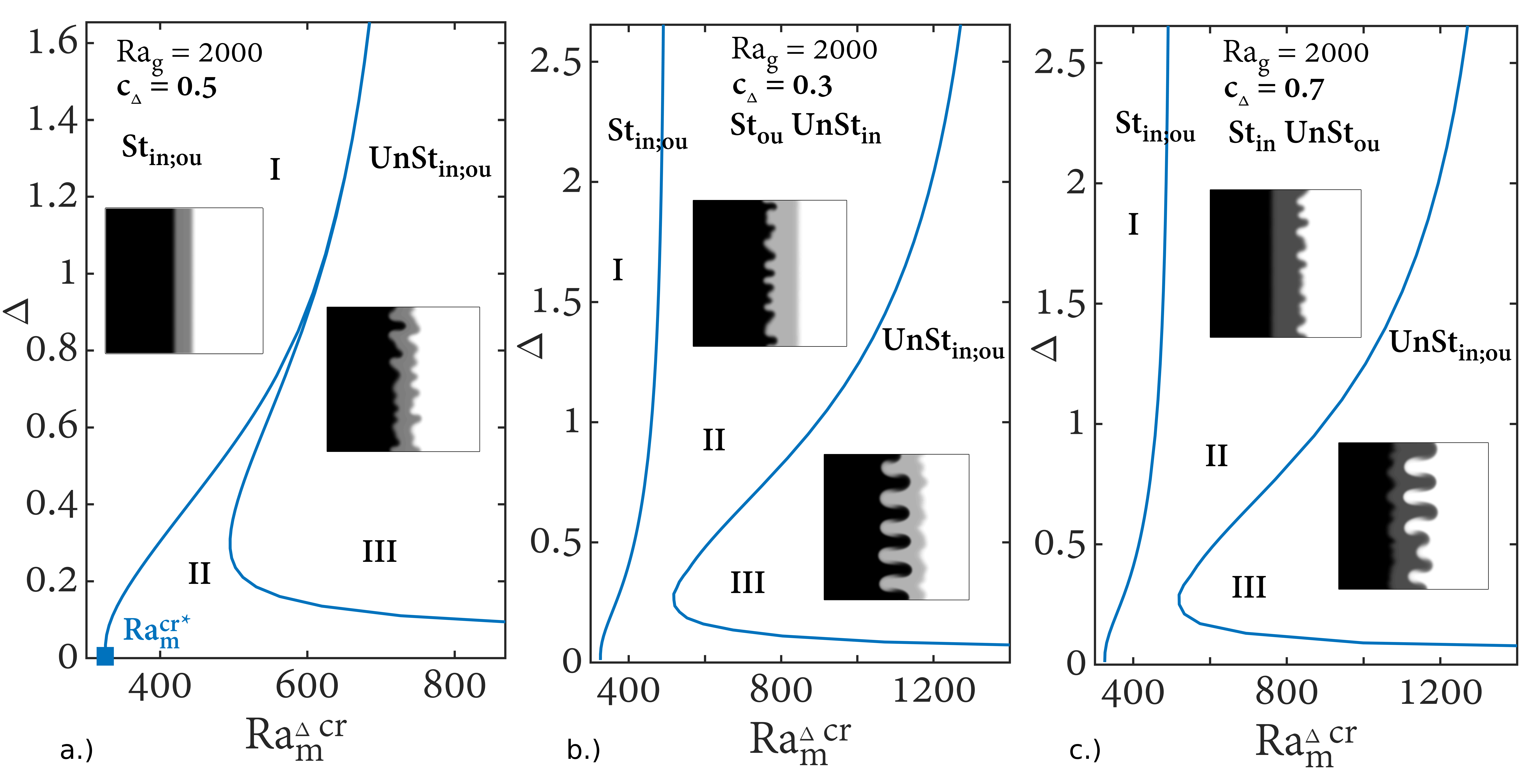}}
\caption{(Colour online) The stability/instability areas of the middle layer from critical magnetic field $\Ram$ over the thickness $\Delta$  for various values of concentration of magnetic particles in middle layer a.) $c_{\Delta} =0.5$; b.) $c_{\Delta} =0.3$ and c.) $c_{\Delta} =0.7$. Data for $\Rag =  2000$.
Roman numerals identify regions in which both boundaries of the mixed layer are stable (I), one boundary is stable and the other unstable (II), or both boundaries are unstable (III). In the boundary labels St and UnSt denote stable and unstable boundaries, respectively; the subscripts in and ou refer to the boundaries with the magnetic fluid and with non-magnetic, and the combined subscript in;ou indicates that the statement applies to both boundaries.
}       
\label{fig:LinAnaliz:StabilIntabilArea}
\end{figure}

Thus if value of magnetic field $\Ram$ is larger than the critical magnetic Rayleigh number $\Ram_{c}$ for two phases of magnetic and nonmagnetic fluids  and less as critical magnetic Rayleigh number $\Ram^{\Delta c}$ for layer of  magnetic fluid between magnetic and nonmagnetic fluids, the initial instability of boundary between two phases of magnetic and non-magnetic fluids $\Ram^{c} < \Ram < \Ram^{\Delta c}$ develop in nonlinear regime and  due to micro-convection and diffusion processes the stable smearing layer is shown in figure~\ref{fig:LinAnaliz:StabilIntabilArea}~(a,b,c) and corresponds to region $\rm I$. 
The linear analysis shows that if the concentration of magnetic particles $c_{\Delta}$ in the middle layer differ from $0.5$  $ c_{\Delta} <0.5$  or $c_{\Delta} > 0.5 $, one boundary between the phases can be unstable and second boundary will be stable, depending from parameters ($\Ram, \Delta, \Rag$ ), indicated in figure~\ref{fig:LinAnaliz:StabilIntabilArea}~(a,b,c)   as region II. 
These results correlate  with numerical simulation data  in figure~\ref{fig:fig2}  and experimental  results in figure~\ref{fig:imgrid_all} where  the second front of  instability  appears on the inner boundary between the  layer and magnetic fluid for some critical values of magnetic field.   
If the value of magnetic field  is more than  critical value of Rayleigh number  $\Ram$ for two boundaries  then both boundaries can be unstable, which in  figure~\ref{fig:LinAnaliz:StabilIntabilArea} (a,b,c) correspond to  region - III. 
In dependence from  concentration of the middle layer the instability  with more  intensity will be dominant on the one of the boundary (for case $c_{\Delta} = 0.5$  and a larger thicknesses of layer $\Delta > 1$ will develop an instability  equal on both boundaries. 
It is worth to note that figure~\ref{fig:LinAnaliz:StabilIntabilArea} can be related to the mastercurve and figure~\ref{fig:mastercurve}, as a clear field dependence is visible.


We also check the fastest growing modes to see how the wavenumber of fingering changes. 
Change in the thicknesses of the middle layer $\Delta$ and magnetic effects $\Ram$ has little influence on characteristic wavenumber, however, with increasing gravitational force, the size of fingers should decrease notably.
For example, an increase in the gravitational Rayleigh number from $\Rag=1000$ till $\Rag=5000$ increases the wavenumber  approximately 2 times from $k \simeq 5$ till $k \simeq 9.5$.
 
\section{\label{sec:concl}Conclusions}
Within this study, magnetic micro-convection was successfully observed experimentally for initially stagnant miscible fluids.
We demonstrated with numerical simulations, linear analysis, and experimentally that gravity limits the micro-convection, by restricting the mixing length with the micro-convective length $\delta_\text{MC}$. 
The experimental results of micro-convective mixing as a relationship between $\delta_\text{MC}$ and $\Ram$ show good quantitative and qualitative agreement both with numerical simulations and with linear analysis for all values of $\Rag$ in this study. 
Moreover, we managed to construct a master curve for this process, which is obtained using the critical magnetic Rayleigh number.
The Micro-convective length $\delta_\text{MC}$ is smaller for higher values of the gravitational Rayleigh number $\Rag$, when the magnetic Rayleigh number is kept constant $\Ram$.

The critical magnetic fields of the instability were calculated theoretically and measured experimentally. 
The results show reasonable agreement with the estimates of the previous study \citep{Kitenbergs_EPJE}, where the differences are justifiable.
Smaller values of $\Ram$ are necessary in order to induce the micro-convective instability in the experiments with smaller values of $\Rag$. 
The effects of the premixed layer on the dynamic of the micro-convection were briefly inspected by numerical simulations. 
The critical values of $\Ram$ and the magnetic fields are accordingly higher if the thickness of the premixed layer is greater. 
This will be investigated experimentally and studied in detail separately.

The theoretical model was improved as a system of three liquid layers with isotropic initial magnetic particle concentration was investigated by the linear analysis. 
This system matches the experimental reality better than the one in the previous study \citep{Kitenbergs_EPJE} with two layers, since experimentally in our setup there is always a premixed layer of both fluids at the beginning of the experiment. 
The dynamics of the micro-convection was calculated by using a Brinkman model and investigating the stability on both borders between two phases of liquids.

\backsection[Acknowledgements]{We thank D. Talbot (PHENIX lab, Sorbonne University in Paris, France) for the magnetic fluid, I. Drikis (MMML lab, University of Latvia) for an Arduino based synchronization solution for microfluidic pumps and M.M.Maiorov (Institute of Physics, University of Latvia) for magnetization measurements.
}

\backsection[Funding]{L.P-S.'s research was supported by the European Social Fund Project No. 8.2.2.0/20/I/006. G.K.'s research has been funded by a PostDocLatvia project No.1.1.1.2/VIAA/1/16/197. }

\backsection[Author contribution statement]{GK and LPS performed the experiments. LPS analyzed the data.  AT updated the theoretical model and performed the numerical simulations and the stability analysis. AC consulted on the theoretical aspects of the study. GK supervised the study. LPS and AT wrote the manuscript. All authors participated in the improvement of
the manuscript. All authors have read and approved the final manuscript. AT and LPS contributed equally to this paper and are co-first authors.}

\appendix

\section{Solution of linear analysis}\label{appA}
Here the details of the linear analysis solution are described. 
The dispersion equation for the growth increment of the instability when $t_{0}\rightarrow 0$ is explored.

The boundary conditions at the discontinuity of the concentration $c_{0}$ are given by the continuity of the concentration perturbation, tangential and normal to the front velocity components and their derivatives:

\begin{equation}
\label{Apend:A:Bound}
\left.
{ \everymath={\displaystyle}
    \begin{array}{cc}
      c_{0}(0^{+}) -  c_{0}(0^{-})= c_{\Delta}-1 \text{;} \\
      c_{0}(\Delta^{+}) -  c_{0}(\Delta^{-})= -c_{\Delta} \text{;} \\
      c^{'}(0^{+}) -  c^{'}(0^{-})= 0\text{;} \\
      c^{'}(\Delta^{+}) -  c^{'}(\Delta^{-})= 0\text{;} \\
      u_x^{'}(0^{+}) -  u_x^{'}(0^{-}) = 0 \text{;} \\
      u_x^{'}(\Delta^{+}) -  u_x^{'}(\Delta^{-}) = 0  \text{;} \\
       \frac{\mathrm{d} c^{'}}{\mathrm{d} x}(0^{+}) -  \frac{\mathrm{d} c^{'}}{\mathrm{d} x}(0^{-})= c_{\Delta} u_x^{'}(0^{+})  - u_x^{'}(0^{-})\text{;} \\
      \displaystyle \frac{\mathrm{d} c^{'}}{\mathrm{d} x}(\Delta^{+}) -  \frac{\mathrm{d} c^{'}}{\mathrm{d} x}(\Delta^{-})= - c_{\Delta}u_x^{'}(\Delta^{-})  \text{;} \\
       \displaystyle \dfrac{\mathrm{d} u_x^{'}}{\mathrm{d} x}(0^{+}) -  \dfrac{\mathrm{d} u_x^{'}}{\mathrm{d} x}(0^{-})   = 0 \text{;} \\  
        \displaystyle { \dfrac{\mathrm{d} u_x^{'}}{\mathrm{d} x}(\Delta^{+}) -  \dfrac{\mathrm{d} u_x^{'}}{\mathrm{d} x}(\Delta^{-})   = 0 } \text{;}  \\
         \displaystyle \dfrac{\mathrm{d}^2 u_x^{'}}{\mathrm{d} x^2}(0^{+}) -  \dfrac{\mathrm{d}^2 u_x^{'}}{\mathrm{d} x^2}(0^{-})   = 0  \text{;} \\  
        \displaystyle { \dfrac{\mathrm{d}^2 u_x^{'}}{\mathrm{d} x^2}(\Delta^{+}) -  \dfrac{\mathrm{d}^2 u_x^{'}}{\mathrm{d} x^2}(\Delta^{-})   = 0} \\
        \displaystyle \dfrac{\mathrm{d}^3 u_x^{'}}{\mathrm{d} x^3}(0^{+}) -  \dfrac{\mathrm{d}^3 u_x^{'}}{\mathrm{d} x^3}(0^{-})   = 24  k^2 (c_{\Delta}-1)\textit{Ra}_{ m} \psi_{\rm m}^{'}(0)  \text{;} \\  
        \displaystyle { \dfrac{\mathrm{d}^3 u_x^{'}}{\mathrm{d} x^3}(\Delta^{+}) -  \dfrac{\mathrm{d}^3 u_x^{'}}{\mathrm{d} x^3}(\Delta^{-})   = -24  k^2 c_{\Delta}\textit{Ra}_{ m} \psi_{\rm m}^{'}(\Delta)}  \text{.}
    \end{array}
    }
  \right \} 
  \end{equation}

The perturbation of the magnetostatic potential and concentration is expressed as:

\begin{eqnarray*}
 \psi_m^{'}(x) = 2\int_{-\infty}^{\infty}c'(\zeta+x)(K_0(k|\zeta|)-K_0(k\sqrt{\zeta^2+1}))\mathrm{ d} \zeta 
  \end{eqnarray*}
and 
 \begin{eqnarray*}
c'(x)|_{x<0} = A \exp(skx) 
\\
c'(x)|_{0<x<\Delta} = C \exp(-skx)  + D \exp(skx) 
\\
c'(x)|_{x>\Delta} = B \exp(-skx) 
  \end{eqnarray*}

Here $K_0$ is the modified Bessel function of the second kind (McDonald function) $ K_0(x) =\displaystyle{ \int_{0}^{\infty}\frac{\cos{(x t )} \rm{d} t}{\sqrt{1+t^2}} } $, parameter s is $s = \sqrt{1+\dfrac{\lambda}{k^2}}$. Equations Eq.(\ref{Eq:Lin:Pert1}) and Eq.(\ref{Eq:Lin:Pert2}) at boundary conditions Eq.(\ref{Apend:A:Bound}) are analysed in the quasi-stationary case. 
The solutions of Eq.(\ref{Eq:Lin:Pert1}) and Eq.(\ref{Eq:Lin:Pert2}) that satisfy the boundary
conditions $ \{ u^{'}(x),\dfrac{\partial u_x^{'}(x)}{\partial x}, c^{'}(x),\psi_m^{'}(x) \}  \rightarrow 0 \quad \text{at} \quad  x \rightarrow \pm\infty $.

For derivative of magneto-static potential take into account $ \{c_0(x,0)|_{-\infty}^{0}= 1; c_0(x,0)|_{0}^{\Delta}=c_{\Delta}, c_0(x,0)|_{\Delta}^{+\infty}=0 \}$ give the extension for interval $(-\infty,+\infty)$

\begin{eqnarray*}
\dfrac{\mathrm{ d} \psi_{\rm m0}(x,0)}{\mathrm{ d} x} = (c_{\Delta}-1)\ln(1+x^{-2})  - c_{\Delta} \ln(1+(x-\Delta)^{-2}) 
  \end{eqnarray*}

The solutions for velocity for three intervals of $x \in$  $[-\infty,0], [0,\Delta]$ and $[\Delta,\infty]$ reads

\begin{eqnarray*}
\tilde{u}^{'}_{x}|_{x<0} = Q_{10} \exp( k x ) + Q_7 \exp( -k x ) +  Q_4 \exp( k m x ) + Q_1 \exp( -k m x ) + 
\\
+ k A \exp( k x ) W(k(s-1),0,x)   - k A \exp( -k x ) W(k(s+1),0,x)-
\\
-\frac{k}{m} A \exp( k m x ) W(k(s-m),0,x)  + \frac{k}{m} A \exp(- k m x ) W(k(s+m),0,x) 
\\
\tilde{u}^{'}_{x}|_{0<x<\Delta} = Q_{11} e^{ k x } + Q_8 e^{ -k x } + Q_5 e^{ km x } + Q_2 e^{ -k m x } +k e^{ k x } \Big [  C   W(-k(s+1),0,x)  + 
\\
+D  W(k(s-1),0,x)    \Big ]  - k \exp( -k x ) \Big [ C   W(-k(s-1),0,x) +D  W(k(s+1),0,x)     \Big ] -
\\
-\frac{k \exp( k m x )}{m} \Big [  C   W(-k(s+m),0,x)  + D   W(k(s-m),0,x)      \Big ] +
\\
+\frac{k \exp( -k m x )}{m} \Big [  C  W(-k(s-m),0,x)  +  D  W(k(s+m),0,x) \Big ]
\\
\tilde{u}^{'}_{x}|_{x>\Delta} = Q_{12} \exp( k x ) + Q_9 \exp( -k x ) + Q_6 \exp( km x ) + Q_3 \exp( -k m x ) + 
\\
+k B  \exp( k x ) W(-k(s+1),\Delta,x)  -k B \exp( -k x ) W(-k(s-1),\Delta,x)  -
\\
-\frac{k}{m} B  \exp( km x ) W(-k(s+m),\Delta,x)  + \frac{k}{m} B  \exp( -k mx ) W(-k(s-m),\Delta,x)
\end{eqnarray*}
the function $W(a,w,z)$ is defined 

\begin{eqnarray*}
  W(a,w,z)=  \int_{w}^{z} \exp(a \zeta) M(c_{\Delta},\Delta,\zeta) {\mathrm{ d}}\zeta
  \\
  M(c_{\Delta},\Delta,x) =  \Ram  \Big( (c_{\Delta}-1)\ln(1+x^{-2})  - c_{\Delta} \ln(1+(x-\Delta)^{-2})  \Big)   + \frac{\Rag}{2}
\end{eqnarray*}

The boundary conditions Eq.(\ref{Apend:A:Bound}) give the set of sixteen equations in matrix form $D  V = 0 $ for the unknown set of constants $ V = {A; B; C; D; Q_1; Q_2; Q_3; Q_4; Q_5; Q_6; Q_7; Q_8; Q_9; Q_{10}; Q_{11}; Q_{12} }$ and solubility condition $\det (D) = 0$ gives the dispersion equation for the growth increment $\lambda$ of small perturbations.

D = \begin{eqnarray*}
\tiny{
\left[ \begin{array}{cccccccccccccccc} 
1 & 0 & 0& 0& 0& 0& 1& 0& 0& 0& 0& 0&d_{1,13}& 0&0&0 \\
0 & 0 & 0& 0& 0& 1& 0& 0& 0& 0& 0&1& 0& d_{2,14}&0&0\\
1&-1&0&1&-1&0&1&-1&0&1&-1&0&0& 0&0&0 \\
0& \varepsilon
^{-m}& -\varepsilon
^{-m}& 0& \varepsilon
^{m}& -e^{m}& 0& \varepsilon
^{-1}& -\varepsilon
^{-1}& 0& \varepsilon
& -\varepsilon
&0& 0&U_2&U_1 \\
-1& c_{\Delta}& 0& -1& c_{\Delta}& 0& -1& c_{\Delta}& 0& -1& c_{\Delta}& 0&sk& 0 &sk&-sk \\
0&c_{\Delta}\varepsilon
^{-m}&0&0&c_{\Delta}\varepsilon
^{m}&0&0&c_{\Delta}\varepsilon
^{-1}&0&0&c_{\Delta}\varepsilon
&0&0 &d_{6,14}&d_{6,15}&d_{6,16} \\
m&-m&0&-m&m&0&1&-1&0&-1&1&0&0&0&0&0 \\
0&m\varepsilon
^{-m}&-m\varepsilon
^{-m}&0&-m\varepsilon
^{m}&m\varepsilon
^{m}&0&\varepsilon
^{-1}&-\varepsilon
^{-1}&0&-\varepsilon
&\varepsilon
& 0 & 0 & U_4 & U_3\\
-m^2& m^2& 0& -m^2& m^2& 0& -1& 1& 0& -1& 1& 0& 0& 0&0&0\\ 
0& -m^2 \varepsilon
^{-m}&m^2\varepsilon
^{-m}&0&-m^2\varepsilon
^{m}&m^2\varepsilon
^{m}&0&-\varepsilon
^{-1}&\varepsilon
^{-1}&0&-\varepsilon
&\varepsilon
&  0 & 0 & U_6 & U_5 \\ 
m^3&-m^3&0&-m^3&m^3&0&1&-1&0&-1&1&0&d_{11,13} & d_{11,14}  &d_{11,15} &d_{11,16} \\  

0& m^3 \varepsilon
^{-m}&-m^3\varepsilon
^{-m}&0&-m^3\varepsilon
^{m}&m^3\varepsilon
^{m}&0&\varepsilon
^{-1}&-\varepsilon
^{-1}&0&-\varepsilon
&\varepsilon
&d_{12,13}& d_{12,14}&d_{12,15}&d_{12,16} \\
-m&0&0&0&0&0&-1&0&0&0&0&0&d_{13,13}& 0&0&0 \\
0&0&0&0&0&m&0&0&0&0&0&1&0&d_{14,14} &0&0 \\
0&0&0&0&0&0&0&0&0&0&0&0 &0 &-\varepsilon^{-s} &\varepsilon^{-s} &\varepsilon^{s} \\
0&0&0&0&0&0&0&0&0&0&0&0 &1 &0 &-1 &-1 \\ 
\end{array} \right ]  }
\end{eqnarray*}

the coefficients $d_{ij}$ and $\varepsilon
$ are defined
 \begin{eqnarray*}
 \varepsilon = \exp(k\Delta)
 \\
d_{1, 13} =  - S_{2}^{\infty} +\frac{S_{2m}^{\infty}}{m} 
\\
d_{2, 14} =   S_{3}^{\Delta}  - \frac{S_{3m}^{\Delta}}{m}   
\\
d_{6, 14} =  -s k  \varepsilon^{-s}   
\\
d_{6, 15} = -c_{\Delta} U_2  +  s k  \varepsilon^{-s}   
\\
d_{6, 16} =    - c_{\Delta} U_1 -  s k \varepsilon^{s}    
\\
d_{11, 13} =   - \frac{48}{k}  \textit{Ra}_{ m} (c_{\Delta} - 1) Y_{0}^{\infty}  
\\
d_{11, 14} =    - \frac{48}{k}  \textit{Ra}_{ m} (c_{\Delta} - 1)  Y_{\Delta}^{\infty} 
\\
d_{11, 15} =   - \frac{48}{k}  \textit{Ra}_{ m} (c_{\Delta} - 1) Y_{0-}^{\Delta}
\\
d_{11, 16} =    - \frac{48}{k}  \textit{Ra}_{ m} (c_{\Delta} - 1)  Y_{0+}^{\Delta}  
\\
d_{12, 13} =   \frac{48}{k} \textit{Ra}_{ m}  c_{\Delta}   J_{0-}^{\infty} 
\\
d_{12, 14} =    \frac{48}{k} \textit{Ra}_{ m}  c_{\Delta}   Y_{\Delta}^{\infty}  
\\
d_{12, 15} =  \frac{48}{k} \textit{Ra}_{ m}  c_{\Delta}   J_{0+}^{\Delta}  -  U_8  
\\
d_{12, 16} =    \frac{48}{k} \textit{Ra}_{ m}  c_{\Delta}    Y_{0+}^{\Delta} -  U_7 
\\
d_{13, 13} = - S_2^{\infty}  +  S_{2m}^{\infty}   
\\
d_{14, 14} = -  S_{3}^{\Delta}  +  S_{3m}^{\Delta} 
\end{eqnarray*}

 corresponding integrals:

\begin{eqnarray*}
U_1 = \Big [-     Q_{1+}^{\Delta} +   Q_{2+}^{\Delta}  +  \frac{1}{m}  \Big (  Q_{1+}^{m\Delta}  - Q_{2+}^{m\Delta}  \Big )   \Big ]
\\
U_2 = [    -     Q_{2-}^{\Delta}  +    Q_{1-}^{\Delta}  +  \frac{1}{m} \Big (  Q_{2-}^{m\Delta}  -    Q_{1-}^{m\Delta} \Big )  \Big ]
\\
U_3 =  \Big [    Q_{1+}^{\Delta}   + Q_{2+}^{\Delta} -   \Big (  Q_{1+}^{m\Delta}  +  Q_{2+}^{m\Delta} \Big )   \Big ]  
\\
U_4 = \Big [    Q_{2-}^{\Delta}  + Q_{1-}^{\Delta}  -    \Big (    Q_{2-}^{m\Delta}  +  Q_{1-}^{m\Delta}   \Big ) \Big ]
\\
U_5 = \Big  [  Q_{1+}^{\Delta}   -  Q_{2+}^{\Delta} -  m \Big (     Q_{1+}^{m\Delta}  -     Q_{2+}^{m\Delta}   \Big ) \Big ]   
\\
U_6  = \Big [ -  Q_{1-}^{\Delta} + Q_{2-}^{\Delta}      -  m \Big (  Q_{2-}^{m\Delta}   -    Q_{1-}^{m\Delta} \Big ) \Big ] 
\\
U_7 = \Big [  - Q_{1+}^{\Delta}   -  Q_{2+}^{\Delta} +  m^2 \Big (  Q_{1+}^{m\Delta}  +   Q_{2+}^{m\Delta}  \Big ) \Big ]  
\\
U_8  = \Big [  -   Q_{2-}^{\Delta}  - Q_{1-}^{\Delta}  +   m^2 \Big (   Q_{2-}^{m\Delta}  + Q_{1-}^{m\Delta} \Big )    \Big ]
\end{eqnarray*}
 
   \begin{eqnarray*}
Q_{1-}^{\Delta} = k\int_{0}^{\Delta} e^{-k(s-1)\zeta-k\Delta} M(c_{\Delta},\Delta,\zeta) d\zeta 
\\
Q_{1+}^{\Delta} = k\int_{0}^{\Delta} e^{k(s-1)\zeta + k\Delta} M(c_{\Delta},\Delta,\zeta) d\zeta  \\
Q_{2-}^{\Delta} = k\int_{0}^{\Delta} e^{-k(s+1)\zeta + k\Delta} M(c_{\Delta},\Delta,\zeta) d\zeta 
\\
Q_{2+}^{\Delta} = k\int_{0}^{\Delta} e^{k(s+1)\zeta - k\Delta} M(c_{\Delta},\Delta,\zeta) d\zeta  
\\
Q_{1-}^{m\Delta} = {k } \int_{0}^{\Delta} e^{-k(s-m)\zeta  - k m \Delta} M(c_{\Delta},\Delta,\zeta) d\zeta
\\
Q_{1+}^{m\Delta} = {k }  \int_{0}^{\Delta} e^{k(s-m)\zeta +  k m \Delta} M(c_{\Delta},\Delta,\zeta) d\zeta  \\
Q_{2-}^{m\Delta} = {k }  \int_{0}^{\Delta} e^{-k(s+m)\zeta + k m \Delta} M(c_{\Delta},\Delta,\zeta) d\zeta
\\
Q_{2+}^{m\Delta} = {k }  \int_{0}^{\Delta} e^{k(s+m)\zeta- k m \Delta} M(c_{\Delta},\Delta,\zeta) d\zeta 
\end{eqnarray*}

      \begin{eqnarray*}
 Y_0^{\infty} = \int_{0}^{\infty}  e^{-sk\zeta} G(\zeta) { \rm d} \zeta  
 \\
  Y_{\Delta}^{\infty} = \int_{\Delta}^{\infty}  e^{-sk\zeta} G(\zeta) { \rm d} \zeta  
 \\
  Y_{0-}^{\Delta} = \int_{0}^{\Delta}  e^{-sk\zeta} G(\zeta) { \rm d} \zeta 
   \\
     Y_{0+}^{\Delta} = \int_{0}^{\Delta}  e^{sk\zeta} G(\zeta) { \rm d} \zeta 
   \\
  J_{0-}^{\Delta} = \int_{0}^{\Delta}  e^{sk(\zeta-\Delta)} G(\zeta) { \rm d} \zeta  
   \\
  J_{\Delta-}^{\infty} = \int_{\Delta}^{\infty}  e^{-sk(\zeta-\Delta)} G(\zeta) { \rm d} \zeta  
     \\
  J_{0-}^{\infty} = \int_{0}^{\infty}  e^{-sk(\zeta-\Delta)} G(\zeta) { \rm d} \zeta  
  \\
    J_{0+}^{\Delta} = \int_{0}^{\Delta}  e^{-sk(\zeta+\Delta)} G(\zeta) { \rm d} \zeta  
   \end{eqnarray*}

 \begin{eqnarray*}
S_2^{\infty} = k\int_{0}^{-\infty} e^{k(s+1)\zeta} M(c_{\Delta},\Delta,\zeta) d\zeta   
\\
S_{2m}^{\infty} = k \int_{0}^{-\infty} e^{k(s+m)\zeta} M(c_{\Delta},\Delta,\zeta) d\zeta   
\\
S_{3}^{\Delta} = k e^{sk\Delta}\int_{\Delta}^{\infty} e^{-k(s+1)\zeta} M(c_{\Delta},\Delta,\zeta) d\zeta   
\\
S_{3m}^{\Delta} = k e^{sk\Delta}\int_{\Delta}^{\infty} e^{-k(s+m)\zeta} M(c_{\Delta},\Delta,\zeta) d\zeta   
\end{eqnarray*}

 functions ${G}(\zeta)$ and ${M}(\zeta)$ are 

  \begin{eqnarray*}
{M}(\zeta) =  \textit{Ra}_{ m}  \Big(  (c_{\Delta}-1)\ln \Big (1+\frac{1}{\zeta^{2}}\Big )  - c_{\Delta} \ln \Big(1+\frac{1}{(\zeta-{\Delta})^{2}} \Big) \Big ) +  \frac{\textit{Ra}_g}{2}  
\end{eqnarray*} 

\begin{eqnarray*}
{G}(\zeta) = K_0(k|\zeta|)-K_0(k\sqrt{\zeta^2+1^2})
  \end{eqnarray*}

\section{Additional measurements of magnetic micro-convection dynamics}\label{appB}
The dynamics of the magnetic micro-convection can be seen in the movies in the Supplementary data. The experiments are recorded on the left side of the movies and the corresponding numerical simulations of micro-convection dynamics is demonstrated on the right side of the movies. The scale within a movie is the same for both an experiment and a numerical simulation. Some visual differences arise between the experiments and the numerical simulations, as the experimental movies are recorded as the light-intensity plots, whereas the numerical simulations are represented as concentration plots. In the movies (1, 2, 3, 4) magnetic fluid $MF_1$ ($\Rag=4657$) is presented at various magnetic fields. In the movie~5 behaviour of a diluted fluid $MF_2$ ($\Rag=3031$) at a high magnetic field $H=106$~Oe ($\Ram=3406$) can be observed. The experiments of movies~4 and 5 have the same magnetic field $H=106$~Oe applied, but the values of the magnetic Rayleigh number are different ($\Ram=7664$ and $\Ram=3406$ accordingly) as the degree of dilution affects it. The more diluted the original magnetic fluid $MF_1$ the smaller the value of $\Ram$ for the same external magnetic field $H$. The diluted fluid $MF_2$ has shorter micro-convective fingers than the concentrated fluid $MF_1$ when the same magnetic field is applied. The movie~6 demonstrates how the thickness of the initial smearing (expressed as $\delta_0$ experimentally and $t_0$ in the simulations) affects the dynamic of the magnetic micro-convection. 

\begin{figure}
 \centerline{\includegraphics[width=0.9\columnwidth]{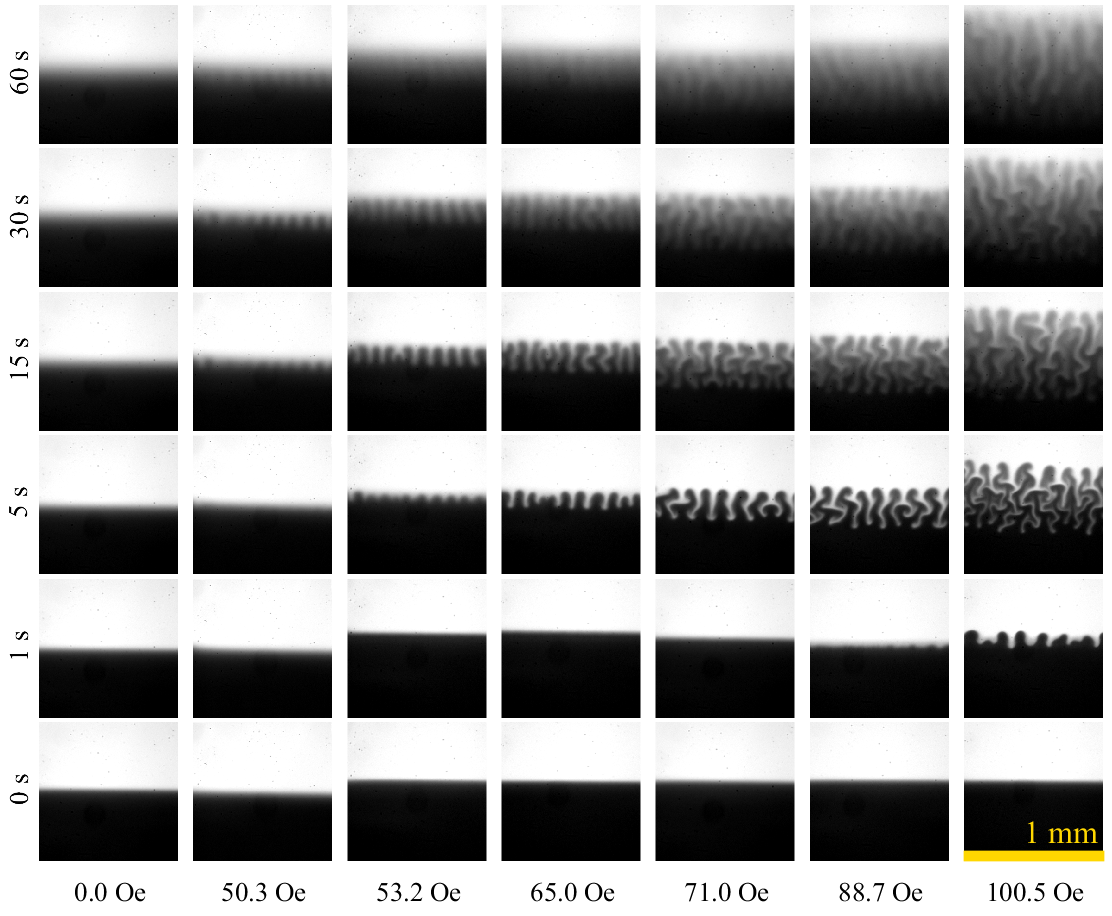}}
\caption{ Image series of magnetic micro-convection dynamics with magnetic fluid with $MF_2$ in various magnetic fields. The contrast of the images is changed, so that pure water would appear white and pure magnetic fluid black. }
\label{fig:imgrid_all_067}      
\end{figure}

\begin{figure}
 \centerline{\includegraphics[width=0.7\columnwidth]{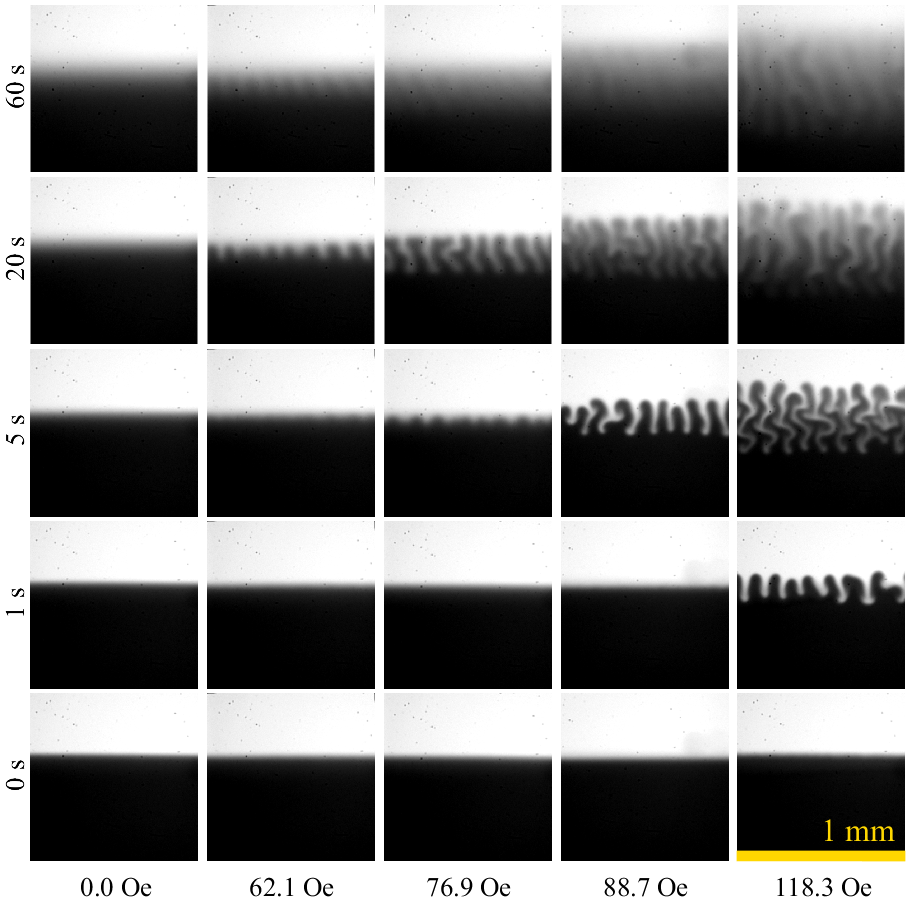}}
\caption{ Image series of magnetic micro-convection dynamics with magnetic fluid with $MF_3$ in various magnetic fields. The contrast of the images is changed, so that pure water would appear white and pure magnetic fluid black.}
\label{fig:imgrid_all_05}      
\end{figure}

\begin{figure}
 \centerline{\includegraphics[width=0.5\columnwidth]{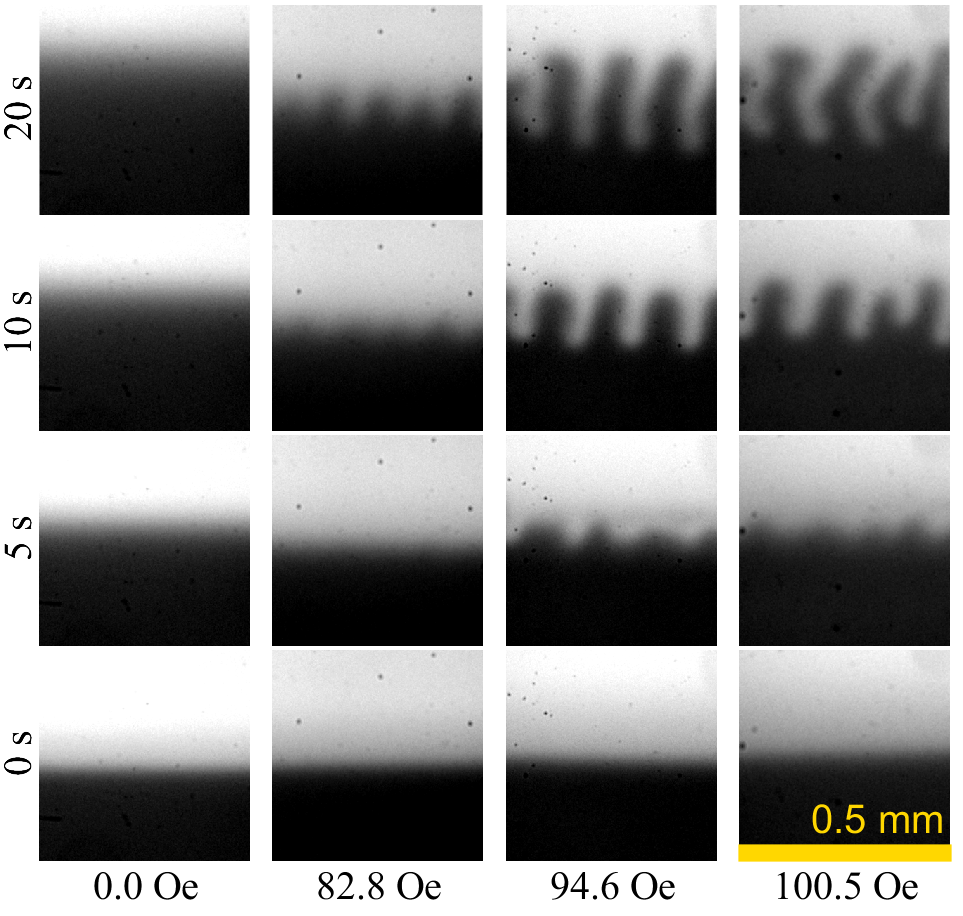}}
\caption{ Image series of magnetic micro-convection dynamics with magnetic fluid with $MF_4$ in various magnetic fields. The contrast of the images is changed, for displaying purpose.}
\label{fig:imgrid_all_033}      
\end{figure}

The dynamics of the magnetic micro-convection for the rest of the magnetic fluids mentioned in this article ($MF_2$, $MF_3$ and $MF_4$ ) are shown in the figures \ref{fig:imgrid_all_067}, \ref{fig:imgrid_all_05} and \ref{fig:imgrid_all_033} respectively. The critical magnetic fields $H_c$ are higher for more diluted magnetic fluids, and the fingers of the instability grow shorter at the same magnetic fields than in $MF_1$. However, the fundamental character of the instability stays the same for all the fluids used. Fingers appear after $H_c$ is reached. At first they grow out straight, but then they bend and branch if the magnetic field is high enough. Once some critical height of the fingers at the specific magnetic field is reached, they do not grow higher, but might still swirl at high magnetic fields. The intense branching of the fingers of the instability could not be reached for $MF_4$ as it would require a magnetic field higher than the limit of our experimental system.
%

\section{Additional measurements of mixing dynamics}\label{appC}
\begin{figure}
 \centerline{\includegraphics[width=0.7\columnwidth]{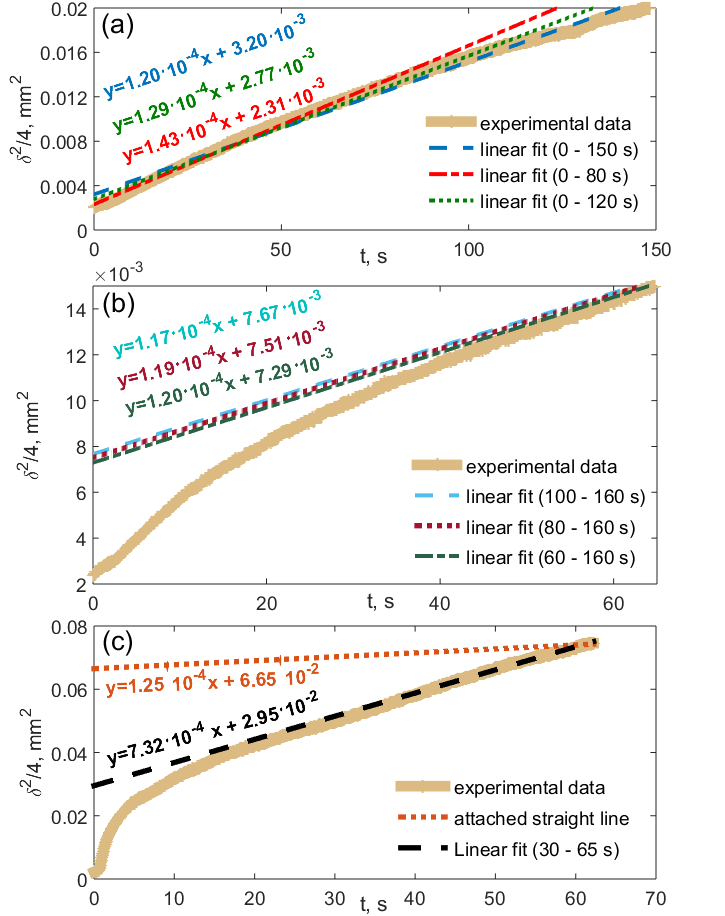}}
\caption{ (Colour online) The mixing between $MF_1$ and water, plotted as $\frac{\delta^2}{4}$ versus time, is shown as a beige line. The diffusion without magnetic field $H=0$~Oe is shown in (a). The blue, green, and red lines are linear fits of the experimental data for different time regions ($150$~s, $120$~s and $80$~s, respectively). (b) demonstrates an experiment at a magnetic field $H=56.2$~Oe and the straight lines are linear fits of the experiment for different time regions. In (c) an experiment ($H=103.5$~Oe) that does not reach the experimental diffusion coefficient $D_\text{exp}$ as a slope is presented. The black dashed line is a linear fit from $t=35$~s to $t=65$~s and the orange dotted line with a slope $D_\text{exp}$ is attached to the graph at $t=65$~s.
}
\label{fig:delta_fitted}      
\end{figure}

As mentioned previously in this article $\delta^2/4$ as a function of time at $H=0$~Oe should be a straight line. As it can be seen in figure~\ref{fig:delta_fitted} (a) the experimental data (beige line) of the mixing of magnetic fluid ($MF_1$ here) when no magnetic field is applied does not look perfectly linear. To approach this fact, the experimental data was linearly fitted with multiple straight lines. The fitting regions were chosen in an according pattern- the first line fitted the data from the start of the experiment (when the pumping of the fluids is stopped at $t=0$~s) until the end of the experiment ($t=250$~s), the next line fitted the experiment from the start, but the fit was $10$~s shorter and so on. The same pattern was repeated with the start times $t=10$~s and $t=25$~s to eliminate some fluctuations of the interface due to stopping the pumps. By calculating the average slope of all these linear fits for all the experiments where $H=0$~Oe for all the magnetic fluids the value of diffusion coefficient $D_\text{exp}=(1.25\pm0.23)\cdot10^{-6}$~cm$^2$/s was found. The average value of $\delta_\text{MC}^2/4$ at $H=0$~Oe is included in the error estimation for $\delta_\text{MC}$, as the value of $\delta_\text{MC}^2/4$ without external magnetic field must be zero, because there is no magnetic micro-convection present, but experimentally there is a rise between the initial smearing of the interface and the y-intercept of the linear fit lines.

Similar approach with several linearly fitted lined were used for the experiments with applied magnetic field (see figure~\ref{fig:delta_fitted} (b)), only here the fit was started from the end of the experiment until approaching some specific time value with $10$~s step.This information was also used to determine the errors of $\delta_\text{MC}$. 

Also mentioned previously in this article, for some experiments, the mixing dynamics between the two fluids, plotted as $\frac{\delta^2}{4}$ versus time, did not reach a linear regime with a slope $D_\text{exp}$. This was observed more frequently for experiments within higher magnetic fields. As shown in figure~\ref{fig:delta_fitted} (c), the linear fit of the experimental data (black dashed line) does not agree with the estimated diffusion coefficient $D_\text{exp}=1.25\cdot10^{-4}$~mm\textsuperscript{2}/s. Although, judging by the eye, the tail of the experimental data (beige line) might seem already linear, it is possible, that a longer experiment would lead to a flatter curve. It was not possible to record longer movies of these experiments, due to heating of the coils (for this particular experiment $H=103.5$~Oe, which is one of the highest magnetic fields at which our experiments were carried out). Therefore, a straight line with a slope of $D_\text{exp}$ was attached at the end of the graph (orange dashed line). That way the $\delta_\text{MC}$ is estimated to be at least the y-intercept of this attached line, while the actual $\delta_\text{MC}$ might be even larger. In this experiment the estimated $\frac{\delta_\text{MC}^2}{4}\geq6.65\cdot10^{-2}$~mm\textsuperscript{2} and accordingly $\delta_\text{MC}\geq0.51$~mm.


\bibliographystyle{jfm}

\bibliography{jfm}





\end{document}